\documentclass[12pt,preprint]{aastex}
\shorttitle{Second order self-gravity calculation on nested grids}
\shortauthors{Wang et al.}
\usepackage{bm}
\usepackage{sansmath}
\usepackage{amsmath}
\usepackage{mathrsfs}
\usepackage{epstopdf}
\usepackage[breaklinks,colorlinks,urlcolor=blue,citecolor=blue,linkcolor=blue]{hyperref}
\usepackage{booktabs}

\newcommand{\hhw}[1]{{\color{black} #1}}

\newcommand{\rt}[1]{{\color{black} #1}}
\newcommand{\hhwang}[1]{{\color{red} #1}}

\begin{document}

\title{Self-Gravitational Force Calculation of Infinitesimally Thin Gaseous Disks on Nested Grids}

\author{Hsiang-Hsu Wang\altaffilmark{1},  Ronald E. Taam\altaffilmark{1,3}, David C. C. Yen \altaffilmark{3}}
\email{yen@math.fju.edu.tw}
\altaffiltext{1}{Institute of Astronomy and Astrophysics, Academia Sinica, P.O. Box 23-141, Taipei 10617, Taiwan, R.O.C.}
\altaffiltext{2}{Department of Mathematics, Fu Jen Catholic University, New Taipei City, Taiwan.}
\altaffiltext{3}{Department of Physics and Astronomy, Northwestern University, 2131 Tech Drive, Evanston, IL 60208, USA}

\begin{abstract}
We extend the work of \citet{Yen2012} and develop 
2nd order formulae to accommodate \hhw{a} nested grid discretization for the direct self-gravitational force calculation 
for infinitesimally thin gaseous disks. This approach uses a two-dimensional kernel derived for infinitesimally thin 
disks and is free of artificial boundary conditions. The self-gravitational force calculation is presented in generalized 
convolution forms for a nested grid configuration.  A numerical technique derived from a fast Fourier transform 
is employed to reduce the computational complexity to be nearly linear. By comparing with analytic potential-density pairs 
associated with the generalized Maclaurin disks, the extended approach is verified to be of second order accuracy using 
numerical simulations. The proposed method is accurate, computationally fast and has the potential to be applied to the 
studies of planetary migration and the gaseous morphology of disk galaxies.  
\end{abstract}


\section{Introduction}
As a direct consequence of the conservation of angular momentum and efficient radiative cooling, thin disks form naturally 
in the Universe. The substructures associated with barred and spiral galaxies, massive stars forming along spiral arms as 
well as the existence of central starburst rings manifest that the self-gravity of gas is important to the evolution of disk 
galaxies \citep{Lin2013, Seo2014, Elmegreen2014, Kim2012, Lee2012, Lee2014}. The self-gravity also plays a role in shaping 
planetary systems during the formation of planets \citep{Inutsuka2010, Zhang2014}. Hydrodynamic simulations including the 
effect of the disk's self-gravity have been used to investigate the orbital evolution of a Jovian planet as reported in 
\citet{Zhang2008}.  The simulations show that the self-gravity of gas in an infinitesimally thin disk plays a significant role 
in the radial drift associated with the type III migration.  The self-gravitational calculation used in~\citet{Zhang2008} is 
based on \hhw{a} uniformly discretized Cartesian grid, and the method has been described in \citet{Yen2012}.  Since the 
self-gravity of gas within the Roche lobe of a protoplanet may influence the planetary migration as well as the mass accretion 
onto the protoplanet, it is highly desirable to perform global simulations with refined grids concentrated around the protoplanet. 

The adequacy of numerical techniques for solving the self-gravity of the gas can be ascertained starting from the gravitational 
potential $\Phi$ associated with the mass density, $\rho$, in three dimensional space, which can be represented by
\begin{eqnarray*}
\nabla \cdot (\nabla \Phi)(x,y,z) = 4\pi G\rho (x,y,z), 
\end{eqnarray*} 
where $\nabla= (\frac{\partial}{\partial x}, \frac{\partial}{\partial y},\frac{\partial}{\partial z})$ or by the volume integral
\begin{eqnarray*}
\label{eqnPhix}
\Phi(x,y,z)=G\int\!\!\!\int\!\!\!\int \frac{-\rho(\bar x,\bar y,\bar z)}{\sqrt{(\bar x- x)^2+(\bar y-y)^2+(\bar z-z)^2}}
d\bar x\, d\bar y\, d\bar z,
\end{eqnarray*}
where $G$ is the gravitational constant.  We are interested in the calculation for an infinitesimally thin disk, where the 
corresponding volumetric mass density $\rho$  is associated with a surface density $\sigma$,
\begin{eqnarray}
\label{eqnrhoxyz}
\rho(x,y,z)=\sigma(x,y)\delta(z),
\end{eqnarray}
where $\delta$ is the Dirac symbol.  Thus, the problem is to solve the potential $\Phi$ induced from 
a surface density, $\sigma$, contained in an infinitesimally thin layer, $\Omega$, satisfying the Poisson equation,
\begin{eqnarray}
\label{eqnDeltaPhi}
\nabla \cdot (\nabla \Phi)(x,y,z)=4\pi G\sigma(x,y) \delta(z), \quad (x,y)\mbox{ in } \Omega.
\end{eqnarray}
The potential in the mid-plane, $\Phi(x,y,0)$, is associated with a kernel integral via 
\begin{eqnarray}
\label{eqnPhixy0G}
\Phi(x,y,0)=G\int\!\!\!\int_\Omega \frac{-\sigma(\bar x,\bar y)}{\sqrt{(x-\bar x)^2+(y-\bar y)^2}} 
d\bar x\,d\bar y.
\end{eqnarray}
For simplicity, we set $G=1$ hereafter.  We note that solving (\ref{eqnDeltaPhi}) is essentially a three-dimensional (3D) 
problem, while the integral form (\ref{eqnPhixy0G}) involves only two-dimensional (2D) calculations when focusing only on 
the forces in the mid-plane. 

Numerous methods have been proposed for three-dimensional potential calculations  
~\citep{Matsumoto2003,Wackers2005,McCorquodale2007,Greengard1987,Anderson1986, Hockney1981,Brandt1977,Smith2004} including the 
Fast Multipole Method (FMM), the Method of Local Corrections (MLC), and FFT-based methods, multigrid, and domain decomposition.  
If one chooses to solve (\ref{eqnDeltaPhi}) using a fast algorithm for the 3D problem, one may reach at best a linear computational 
complexity of $O(N^3)$, where $N$ is the number of zones in one direction. In contrast, solving the integral form (\ref{eqnPhixy0G}) 
may have a computational complexity of only $O(N^2)$. In other words, solving (\ref{eqnPhixy0G}) can be more computationally 
economical than solving (\ref{eqnDeltaPhi}). However, it seems more straightforward to develop a numerical method for a nested 
grid configuration for the differential form (\ref{eqnDeltaPhi}) than that for the integral form (\ref{eqnPhixy0G}), since the 
differentiation operator involves only the local information.  \hhw{We note that the multigrid relaxation method} \rt{is fast, 
flexible and has} \hhw{been used extensively when mesh refinements are required \citep{Hockney1981}. However, the multigrid 
methods, which are by nature only for three-dimensional problems, cannot be reduced \hhwang{to} two-dimensional calculations for an 
infinitesimally thin disk as discussed in this paper. }

A few methods in the literature can be applied to solve the potential induced from an infinitesimally thin disk.  The direct 
$N$-body method is conceptionally simple and relatively straightforward to implement, but it has only first order accuracy and 
high numerical complexity.  The fast Fourier \hhw{transform based} methods have better numerical complexity, but are subject to 
periodic or isolated boundary conditions \citep{James1977}.  \hhw{Its application is restricted to a calculation domain that is 
uniformly} \rt{discretized.} The spectral methods are popular and can solve (\ref{eqnDeltaPhi}) with better accuracy, however, 
artificial boundary conditions need to be specified before the calculations.  The development of this subject has been recently 
reviewed by \citet{Shen2009}. In contrast to those aforementioned works, a direct method for gravitational force calculation has 
been developed in \citet{Yen2012}.  The method has a numerical complexity of $O(N^2 \log (N))$, with a numerical accuracy of 
second order and without the requirement of artificial boundary conditions.  The major objective of this work is to develop the 
formulae and generalize the work of \citet{Yen2012} to accommodate the nested grid configuration. 

The rest of this paper is organized as follows.  In Section 2, we describe the proposed method using a direct integral via the 
Green's function method.  The main concept of the numerical calculations is to recast the calculation of (\ref{eqnPhixy0G}) into 
a generalized convolution form.  In Section 3, a few examples with analytic solutions generalized from \cite{Schulz2009} are 
adopted to verify the order of accuracy and the performance of the method.  We discuss and conclude this work in the last section.

\section{Method}
\hhw{The central idea} \rt{underlying our approach} \hhw{is to solve for the gravitational force, rather than the potential, by 
taking a derivative of the integral expression for} \rt{the} \hhw{potential, which is an integral over the surface density 
convoluted by the (3D) Green's function of the Laplacian.  As this integral would, in general, be difficult to calculate, 
approximations are required} \rt{for} \hhwang{simplification.} 
\hhw{The approximations rely on a discretised domain, and a truncated Taylor expansion of the surface density.  For instance, 
the surface density is approximated by a linear function of $(x, y)$} \rt{within} \hhw{a cell and the approximation follows that 
the accuracy} \rt{is} \hhw{of second order.  The integrals (in $x$ and $y$) can be evaluated within each cell using integrals 
of closed form and the force within a cell} \rt{can} \hhw{be calculated.  These forces can then be summed over cells of all 
patches and all levels in order to find the total forces.  The numerical complexity of a direct calculation is $O(N^4)$, where 
$N$ is the number of} \rt{cells} \hhw{in one dimension.  These forces can be represented} \rt{in} \hhw{a discretized convolution 
form and the complexity is linear and reduced to $O(N^2)$ with the help of FFT.  In general, FFTs in a non-uniform grid present 
difficulties.  In order to have the nested grid formulation be amenable to a FFT approach, one must determine the convolution 
forms for cells in differing patches, which is done in this work.}

The proposed method has two parts. In the first part, we develop the formulae that can be used to approximate the integral 
(\ref{eqnPhixy0G}) with second order accuracy. In particular, these formulae are expressed in a generalized convolution form, 
which can be applied to nested grid configurations.  We note that using the fast Fourier transform for a non-uniformly discretized 
calculation domain is not straightforward. However, we demonstrate in the second part, how the fast Fourier transform can 
be applied to nested grid structures, so that the numerical complexity remains as $O(N^2\ln N)$ and the use of artificial boundary 
conditions is avoided. 

\subsection{Approximation of second order accuracy}
Define the nested  domains $\Omega^{\ell}_k=[L^\ell_k,R^\ell_k]\times
[B^\ell_k,T^\ell_k]$ for $\ell=\ell_{\min},\ell_{\min}+1,\ldots,\ell_{\max}$
 and $k=1,2,\ldots,K^\ell$, where $\ell_{\min}$, $\ell_{\max}$ and $K^\ell$ are positive integers. Here, 
$\ell$ denotes the level of grid and $K^{\ell}$ represents the number of patch for a given level $\ell$. 
That is, there are $\ell_{\max}-\ell_{\min}+1$ grid levels and $K^\ell$ patches for the grid level $\ell$.
The domains $\Omega^\ell_k$ for  $\ell=\ell_{\min},\ell_{\min}+1,\ldots,\ell_{\max}$
 and $k=1,2,\ldots,K^\ell$ are non-overlapping except at their boundaries.  
The \hhw{cell} size used to discretize the domain $\Omega^\ell_k$ is uniform and described by
$(\frac{R^\ell_k-L^\ell_k}{N^\ell_k-1})\times (\frac{T^\ell_k-B^\ell_k}{M^\ell_k-1}):=\Delta x_\ell\times \Delta y_\ell$,
where $N^\ell_k-1$ and $M^\ell_k-1$ are positive integers, corresponding to the number of cells in $x$ and $y$ directions, respectively.
We note that all the patches of the same grid level $\ell$ share the same \hhw{cell} size.  
We further denote the cells of $\Omega^{\ell}_k$ as 
$\Omega^{\ell}_{k,i,j}=\{(x,y): x^\ell_{k,i}\le x \le x^\ell_{k,i+1}, y^\ell_{k,j}\le y \le y^\ell_{k,j+1}\}$,
where $i=0,1,\ldots,N^{\ell}_k-1$
and $j=0,1,\ldots,M^{\ell}_k-1$.
Here, $x^\ell_{k,i}=i (R^\ell_k-L^\ell_k)/N^\ell_k + L^\ell_k=i(\Delta x)^\ell +L^\ell_k$
and   $y^\ell_{k,j}=j (T^\ell_k-B^\ell_k)/M^\ell_k + B^\ell_k=j(\Delta y)^\ell + B^\ell_k$.
The computational domain $\Omega$ is the set 
$\displaystyle \bigcup^{\ell_{\max}}_{\ell=\ell_{\min}}\bigcup^{K^\ell}_{k=0} \Omega^\ell_k$.
Since the interiors of patches $\Omega^{\ell}_k$ are mutually exclusive, except at the boundaries of 
the patches, every point in the calculation domain $\Omega$ can only belong to one specific patch. 

By (\ref{eqnPhixy0G}), the potential induced from the surface density $\sigma$ is rewritten as
\begin{eqnarray}
\label{eqnPhixy0}
\Phi(x,y,0)=\int\!\!\!\int K(\bar x-x,\bar y-y,0)\sigma(\bar x, \bar y)d\bar x d\bar y,
\end{eqnarray}
where $K(x,y,z)=\frac{-1}{\sqrt{x^2+y^2+z^2}}$
is the Green's function of the Laplacian equation in the entire three dimension space.
The $x$-force is associated with the partial derivative in the $x$-direction of the potential defined in (\ref{eqnPhixy0}),
\begin{eqnarray}
\label{eqndPhi}
-\frac{\partial}{\partial x}
\Phi(x,y,0)
&=& -\int\!\!\!\int 
\frac{\partial}{\partial x}
K(\bar x-x,\bar y-y,0) \sigma(\bar x, \bar y) d\bar x d\bar y\nonumber \\
&=& 
\sum^{\ell_{\max}}_{\ell=\ell_{\min}} \sum^{K^\ell}_{k=1}
 -\int\!\!\!\int_{\Omega^{\ell}_k}
\frac{\partial}{\partial x}K(\bar x-x,\bar y-y,0) 
\sigma(\bar x, \bar y) d\bar x d\bar y.
\end{eqnarray}
where $\frac{\partial}{\partial x}K(x,y,z)=\frac{x}{(x^2+y^2+z^2)^{3/2}}$ 
is the partial derivative in $x$-direction of the Green's function $K$. 
Assuming that the surface density $\sigma$ contained in the bounded domain $\Omega$ is sufficiently smooth,
the Taylor expansion of $\sigma(\bar x,\bar y)$
on $\Omega^{\ell}_{k,i,j}$ is represented as
\begin{eqnarray}
\label{eqnSigma}
\sigma(\bar x,\bar y)
\simeq 
\sigma^{\ell}_{k,i,j}
+\delta^{\ell,x}_{k,i,j}(\bar x-x^{\ell}_{k,i+1/2})
+\delta^{\ell,y}_{k,i,j}(\bar y-y^{\ell}_{k,j+1/2}),
\end{eqnarray}
where $\sigma^{\ell}_{k,i,j}$, $\delta^{\ell,x}_{k,i,j}$ and $\delta^{\ell,y}_{k,i,j}$ are constants in the cell $\Omega^{\ell}_{k,i,j}$. 
Substituting (\ref{eqnSigma}) into (\ref{eqndPhi}), 
the $x$-force can be approximated by
\begin{eqnarray*}
-\frac{\partial}{\partial x}
\Phi(x,y,0)
&\simeq&
\sum^{\ell_{\max}}_{\ell=\ell_{\min}} \sum^{K^\ell}_{k=1}
\sum^{N^{\ell}_k-1}_{i=0} \sum^{M^{\ell}_k-1}_{j=0}
 -\int\!\!\!\int_{\Omega^{\ell}_{k,i,j}}
\frac{\partial}{\partial x}K(\bar x-x,\bar y-y,0)\times \\
&&
\left[
\sigma^{\ell}_{k,i,j}
+\delta^{\ell,x}_{k,i,j}(\bar x-x^{\ell}_{k,i+1/2})
+\delta^{\ell,y}_{k,i,j}(\bar y-y^{\ell}_{k,j+1/2})
\right]
d\bar x d\bar y.\\
\end{eqnarray*}
Furthermore, the $x$-force at the cell center 
$(x,y)=(x^{\hat \ell}_{\hat k, \hat i+1/2},y^{\hat \ell}_{\hat k,\hat j+1/2})$ 
of $\Omega^{\hat\ell}_{\hat k, \hat i, \hat j}$ is
\begin{eqnarray*}
-\frac{\partial}{\partial x}
\Phi(x^{\hat \ell}_{\hat k, \hat i+1/2},y^{\hat \ell}_{\hat k, \hat j+1/2},0)
&\simeq &
\sum^{\ell_{\max}}_{\ell=\ell_{\min}} \sum^{K^\ell}_{k=1}
\sum^{i^{\ell}_k}_{i=0} \sum^{j^{\ell}_k}_{j=0}
 -\int\!\!\!\int_{\Omega^{\ell}_{k,i,j}}
\frac{\partial}{\partial x}K(\bar x-x^{\hat \ell}_{\hat k, \hat i+1/2},\bar y-y^{\hat \ell}_{\hat k,\hat j+1/2},0)\times\\
&&
\left[
\sigma^{\ell}_{k,i,j}
+\delta^{\ell,x}_{k,i,j}(\bar x-x^{\hat \ell}_{\hat k,\hat i+1/2})
+\delta^{\ell,y}_{k,i,j}(\bar y-y^{\hat \ell}_{\hat k,\hat j+1/2})
\right]
d\bar x d\bar y\\
&\equiv & F^{x,0}+F^{x,x}+F^{x,y}.\\
\end{eqnarray*}
Thus,
\begin{eqnarray*}
F^{x,0}
&\equiv& 
-\sum^{\ell_{\max}}_{\ell=\ell_{\min}} \sum^{K^\ell}_{k=1}
\sum^{N^{\ell}_k-1}_{i=0} \sum^{M^{\ell}_k-1}_{j=0}
\sigma^{\ell}_{k,i,j}
 \int\!\!\!\int_{\Omega^{\ell}_{k,i,j}}
 \frac{\partial}{\partial x}K(\bar x-x^{\hat \ell}_{\hat k, \hat i+1/2},\bar y-y^{\hat \ell}_{\hat k,\hat j+1/2},0)
 d\bar x d\bar y\\
 &=& 
-\sum^{\ell_{\max}}_{\ell=\ell_{\min}} \sum^{K^\ell}_{k=1}
\sum^{N^{\ell}_k-1}_{i=0} \sum^{M^{\ell}_k-1}_{j=0}
\sigma^{\ell}_{k,i,j}
 \int^{y^{\ell}_{k,j+1}}_{y^{\ell}_{k,j}}
 \!\!\!\int^{x^{\ell}_{k,i+1}}_{x^{\ell}_{k,i}}
 \frac{\partial}{\partial x}K(\bar x-x^{\hat \ell}_{\hat k, \hat i+1/2},\bar y-y^{\hat \ell}_{\hat k,\hat j+1/2},0)
 d\bar x d\bar y\\
  &=& 
\sum^{\ell_{\max}}_{\ell=\ell_{\min}} \sum^{K^\ell}_{k=1}
\sum^{N^{\ell}_k-1}_{i=0} \sum^{M^{\ell}_k-1}_{j=0}
\sigma^{\ell}_{k,i,j}\times {\cal K}^{x,0,\ell,\hat\ell} (\bar x-x^{\hat\ell}_{\hat k,\hat i+1/2},\>
\bar y-y^{\hat\ell}_{\hat k,\hat j+1/2})|^{x^{\ell}_{k,i+1}}_{x^{\ell}_{k,i}}|^{y^{\ell}_{k,j+1}}_{y^{\ell}_{k,j}}.
\end{eqnarray*}
where $g(x)|^b_a=g(b)-g(a)$, and the corresponding antiderivative of $\frac{\partial}{\partial x}K(x,y,0)$ is given by
\begin{eqnarray}
\label{eqnCalK}
{\cal K}^{x,0,\ell,\hat\ell} (x,y)
=-|\mbox{sgn}(x)|\ln(y+\sqrt{x^2+y^2}) 
-(1-|\mbox{sgn}(x)|)\mbox{sgn}(y) \ln (|y|).
\end{eqnarray}
The equation (\ref{eqnCalK}) is derived from 
\begin{eqnarray*}
\int^d_c\int^b_a \frac{x}{(x^2+y^2)^{3/2}} dx\,dy
=\int^d_c -\frac{1}{\sqrt{x^2+y^2}}|^b_a dy.
\end{eqnarray*}
For a nested grid calculation, $a$ or $b$ can be zero. 
In the case that $ab\not=0$,
\begin{eqnarray}
\label{eqnbnot0}
\int^d_c -\frac{1}{\sqrt{b^2+y^2}} dy = -\ln (y+\sqrt{b^2+y^2})|^d_c,
\end{eqnarray}
otherwise, 
\begin{eqnarray}
\label{eqnbeq0}
\int^d_c -\frac{1}{|y|} dy = -\mbox{sgn}(y) \ln |y||^d_c.
\end{eqnarray}
Combining (\ref{eqnbnot0}) and (\ref{eqnbeq0}), (\ref{eqnCalK}) follows.
Compared to Equation (3.11) in~\cite{Yen2012}, (\ref{eqnCalK}) is more general and suitable for 
a nested grid calculation.  
Similarly for $F^{x,x}$ and $F^{x,y}$, which are defined as 
\begin{eqnarray*}
F^{x,x}&=&
\sum^{\ell_{\max}}_{\ell=\ell_{\min}} \sum^{K^\ell}_{k=1}
\sum^{N^{\ell}_k-1}_{i=0} \sum^{M^{\ell}_k-1}_{j=0}
\delta^{\ell,x}_{k,i,j}\times 
{\cal K}^{x,x,\ell,\hat\ell}
(\bar x-x^{\hat\ell}_{\hat k,\hat i+1/2},\>
\bar y-y^{\hat\ell}_{\hat k,\hat j+1/2})|^{x^{\ell}_{k,i+1}}_{x^{\ell}_{k,i}}|^{y^{\ell}_{k,j+1}}_{y^{\ell}_{k,j}},\\
F^{x,y}&=&
\sum^{\ell_{\max}}_{\ell=\ell_{\min}} \sum^{K^\ell}_{k=1}
\sum^{N^{\ell}_k-1}_{i=0} \sum^{M^{\ell}_k-1}_{j=0}
\delta^{\ell,y}_{k,i,j}\times {\cal K}^{x,y,\ell,\hat\ell}
(\bar x-x^{\hat\ell}_{\hat k,\hat i+1/2},\>
\bar y-y^{\hat\ell}_{\hat k,\hat j+1/2})|^{x^{\ell}_{k,i+1}}_{x^{\ell}_{k,i}}|^{y^{\ell}_{k,j+1}}_{y^{\ell}_{k,j}}.
\end{eqnarray*}
Here, 
\begin{eqnarray*}
{\cal K}^{x,x,\ell,\hat\ell}(x,y)
&=&x {\cal K}^{x,0,\ell,\hat\ell}(x,y) - y \ln (x+\sqrt{x^2+y^2})\\
{\cal K}^{x,y,\ell,\hat\ell}(x,y)
&=&y {\cal K}^{x,0,\ell,\hat\ell}(x,y)-\sqrt{x^2+y^2}.
\end{eqnarray*}
We now show that the ${\cal K}^{x,\cdot,\ell,\hat\ell}$ 
can be expressed in a generalized convolution form. 
Let us consider the representation of $x^{\ell}_{k,i}-x^{\hat\ell}_{\hat k,\hat i+1/2}$ in 
\begin{eqnarray*}
{\cal K}^{x,x,\ell,\hat\ell}
(\bar x-x^{\hat\ell}_{\hat k,\hat i+1/2},\>
\bar y-y^{\hat\ell}_{\hat k,\hat j+1/2})|^{x^{\ell}_{k,i+1}}_{x^{\ell}_{k,i}}|^{y^{\ell}_{k,j+1}}_{y^{\ell}_{k,j}}.
\end{eqnarray*} 
This calculation involves the following three cases: 
\begin{enumerate}
\item[Case 1.] When $\ell=\hat \ell$, the term $x^{\ell}_{k,i}-x^{\hat\ell}_{\hat k,\hat i+1/2}$ 
is equal to $\Delta x_\ell (i-\hat i - 1/2) + (L^\ell_k - L^{\hat\ell}_{\hat k})$. 
\item[Case 2.] When $\ell>\hat \ell$, the term 
$x^{\ell}_{k,i}-x^{\hat\ell}_{\hat k,\hat i+1/2}$ 
is equal to $\Delta x_\ell ( i-\frac{\Delta x_{\hat \ell}}{\Delta x_\ell}(\hat i+1/2) ) + (L^\ell_k - L^{\hat\ell}_{\hat k})$.
\item[Case 3.] When $\ell<\hat \ell$, the term 
$x^{\ell}_{k,i}-x^{\hat\ell}_{\hat k,\hat i+1/2}$ 
is equal to $\Delta x_{\hat \ell} ( \frac{\Delta x_\ell}{\Delta x_{\hat\ell}}i-\hat i-1/2 ) + (L^\ell_k - L^{\hat\ell}_{\hat k})$.
\end{enumerate}
Now, we impose the condition that $\Delta x_{\hat\ell}/\Delta x_\ell$ is a positive integer for \hhw{$\ell>\hat\ell$}
and $\Delta x_\ell/\Delta x_{\hat\ell}$ a positive integer for \hhw{$\hat\ell>\ell$}.
We note that in Case 2, the $x^{\ell}_{k,i}$ can be identical to $x^{\hat\ell}_{\hat k,\hat i+1/2}$.
Similar discussion can be also applied to $y^{\ell}_{k,j}-y^{\hat\ell}_{\hat k,\hat j+1/2}$.
In general, the calculation of $F^{x,\cdot}$ involves the following convolution forms,
\hhw{
\begin{eqnarray*}
a_{\hat{n},\hat{p}}&=&\sum_{n}\sum_{p} b_{{\hat n}-n,{\hat p}-p}c_{n,p}, \quad {\rm for~Case~1},\\
a_{\hat{n},\hat{p}}&=&\sum_{n}\sum_{p} b_{m{\hat n}-n,q{\hat p}-p}c_{n,p}, \quad {\rm for~Case~2}, \\
a_{\hat{n},\hat{p}}&=&\sum_{n}\sum_{p} b_{{\hat n}-mn,{\hat p}-qp}c_{n,p}, \quad {\rm for~Case~3},
\end{eqnarray*} 
}
where $m\equiv \frac{\Delta x_{\hat \ell}}{\Delta x_\ell}$, $q\equiv \frac{\Delta y_{\hat \ell}}{\Delta y_\ell}$ for $\ell > {\hat \ell}$ and $m\equiv \frac{\Delta x_{\ell}}{\Delta {x_{\hat \ell}}}$, $q\equiv \frac{\Delta y_{\ell}}{\Delta {y_{\hat \ell}}}$ for $\ell < {\hat \ell}$. The matrix $b$ corresponds to the force kernels, while the matrix $c$ corresponds to the surface density. 
\subsection{Fast calculation for a generalized convolution form}

In this subsection, we calculate the generalized convolution forms in Cases 2 and 3 discussed in the 
previous subsection. 
For the sake of simplicity and clarity, only a calculation in one-dimension is demonstrated. Let 
$\{a_{{n}}\}$, $\{b_{{n}}\}$ and $\{c_{{n}}\}$ be three sequences 
and define the generalized $z$-transform to be $a(z;m,k)=\sum_{{n}} a_{m{n}+k} z^{m{n}+k}$,
where $m$ is a positive integer and $k$ is a non-negative integer.
A similar definition of the $z$-transform is also applied for $b(z;m,k)$ and $c(z;m,k)$.  
The calculation for a one-dimensional convolution problem is the following. 
\begin{enumerate}
\item[Case 1'.]$\ell=\hat\ell$.
In this case, the calculation of gravitational forces is known to be a normal convolution form which can be 
computed using a fast Fourier transform (see \cite{Yen2012}).
\item[Case 2'.]$\ell>\hat \ell$.
 The one-dimensional convolution is generalized to 
\begin{eqnarray}
\label{eqnbmnn}
a_n  = \sum_{\hat n} b_{mn-\hat n}c_{\hat n},\quad n\in {\cal Z},
\end{eqnarray}
where $m>1$ is an integer.
Now,
\begin{eqnarray*}
a(z^m;1,0)&=&
\sum_n a_n z^{mn}  = \sum_n
\sum^{m-1}_{k=0}\sum_{n'} b_{mn-mn'-k}c_{mn'+k}z^{mn}\\
&=& \sum^{m-1}_{k=0}
\sum_{n'}
c_{mn'+k}z^{mn'+k}
\sum_{n}
 b_{m(n-n')-k}z^{m(n-n')-k}\\
&=& \sum^{m-1}_{k=0}b(z;m,-k)c(z;m,k)
\end{eqnarray*}
In this case, the coefficients $\{a_n\}$ are a summation of the products of
the transforms of $\{b_{mn'+k}\}$ and $\{c_{mn'+k}\}$ from $k=0$ up to $k=m-1$.
By applying the fast Fourier transform, the computational complexity is linearly proportional to the lengths of $\{b_n\}$ and $\{c_n\}$.
This one-dimensional calculation corresponds to computing the two-dimensional force 
at the center of $\Omega^{\hat\ell}_{\hat k,\hat i, \hat j}$
contributed from the surface density on a finer cell denoted by $\Omega^\ell_{k,i,j}$.

\item[Case 3'.]$\ell<\hat\ell$.
The one-dimensional convolution is generalized to
\begin{eqnarray}
\label{eqnbnmn}
a_n  = \sum_{\hat n} b_{n-m\hat n}c_{\hat n},\quad n\in {\cal Z},
\end{eqnarray}
where $m>1$ is an integer. Multiplying $z^n$ from both sides of (\ref{eqnbnmn}),
it is
\begin{eqnarray*}
a_n z^{n}
 =\sum_{\hat n} b_{n-m\hat n} c_{\hat n}z^{n}, 
\end{eqnarray*}
and to rewrite $n=mn'+k$, $k=0,1,\ldots,m-1$ and $n'\in {\cal Z}$. 
For a given $k$, where $k=0,1,\ldots, m-1$, the coefficients $\{a_{mn'+k}:n'\in {\cal Z}\}$ can be calculated by  
\begin{eqnarray*}
\sum_{n'} a_{mn'+k}z^{mn'+k}
=\sum_{\hat n}c_{\hat n}z^{m\hat n}
 \sum_{n'} b_{m(n' -\hat n)+k} 
 z^{m(n'-\hat n)+k}
 =c(z^m;1,0) b(z;m,k).
\end{eqnarray*}
In this case, since the sequence $\{a_n\}$ is split into $m$ groups, the calculation of the coefficients $\{a_n\}$ requires $m$ times Fourier transform.
This one-dimensional calculation corresponds to computing the two-dimensional force 
at the center of $\Omega^{\hat\ell}_{\hat k,\hat i,\hat j}$
contributed from the surface density on a coarser cell denoted by $\Omega^\ell_{k,i,j}$.

\end{enumerate}

We demonstrate an example for Case 2' and Case 3'. 
Let two vectors $b$ and $c$ be
\begin{eqnarray*}
b=(b_0,b_1,b_2,b_3,b_4,b_5,b_6,b_7), \quad
c=(c_0,c_1,c_2,c_3).
\end{eqnarray*}
For Case 2' and $m=2$, the vector $a$ defined in (\ref{eqnbmnn}) is
\begin{eqnarray*}
\begin{array}{ll}
a_0 & =b_0 c_0\\
a_1 & =b_0 c_2 + b_1 c_1+ b_2 c_0\\
a_2 & =          b_1 c_3+ b_2 c_2 + b_3 c_1 + b_4 c_0\\
a_3 & =                             b_3 c_3 + b_4 c_2 + b_5 c_1 + b_6 c_0\\
a_4 & =                                                 b_5 c_3 + b_6 c_2 + b_7 c_1\\
a_5 & =                                                                     b_7 c_3\\
\end{array}
\end{eqnarray*}
The generalized $z$-transform of the vector $a$ is 
$a_0 +a_1 z^2+a_2 z^4 + a_3 z^6 +a_4 z^8 + a_5 z^{10}$, which is identical to 
$(b_0+b_2 z^2+b_4 z^4+b_6 z^6)(c_0 + c_2 z^2)+ (b_1z+b_3 z^3 + b_5 z^5 + b_7 z^7)(c_1z+c_3z^3)$.

For Case 3' and $m=2$, the vector $a$ defined in (\ref{eqnbnmn}) is 
\begin{eqnarray*}
\begin{array}{llll}
a_0   &=  b_0 c_0                       & a_1   &=b_1c_0                     \\
a_2   &=  b_0 c_1 + b_2c_0              & a_3   &=b_1c_1+b_3c_0              \\
a_4   &=  b_0 c_2 + b_2c_1+b_4c_0       & a_5   &=b_1c_2+b_3c_1+b_5c_0       \\
a_6   &=  b_0 c_3 + b_2c_2+b_4c_1+b_6c_0& a_7   &=b_1c_3+b_3c_2+b_5c_1+b_7c_0\\
a_8   &=            b_2c_3+b_4c_2+b_6c_1& a_9   &=       b_3c_3+b_5c_2+b_7c_1\\
a_{10}&=                   b_4c_3+b_6c_2& a_{11}&=              b_5c_3+b_7c_2\\
a_{12}&=                          b_6c_3& a_{13}&=                     b_7c_3\\
\end{array}
\end{eqnarray*}
Due to $m=2$, the coefficients of the coefficients of the vector $a$ are split into two groups
$(a_0,a_2,a_4,a_6,a_8,a_{10},a_{12})$ and $(a_1,a_3,a_5,a_7,a_9,a_{11},a_{13})$, which can be  calculated from
the generalized $z$-transform 
\begin{eqnarray*}
(b_0+b_2z^2+b_4z^4+b_6z^6)(c_0+c_1z^2+c_2z^4+c_3z^6),\mbox{ and } 
(b_1z+b_3z^3+b_5z^5+b_7z^7)(c_0+c_1z^2+c_2z^4+c_3z^6),
\end{eqnarray*}
respectively, and compared the results with the generalized $z$-transform of 
$(a_{2k})$ and $(a_{2k+1})$, where $k=0,1,2,\ldots,6$.

\section{Results}
Potential-density pairs of infinitesimally thin disks are adopted to explore the performance of the proposed method. The surface density, which is generalized from the disk model discussed in \cite{Schulz2009}, is described by 
\begin{eqnarray}
\sigma_{D_n}(r;\alpha)=\left\{
\begin{array}{ll}
\sigma_0 (1-\frac{r^2}{\alpha^2})^{n-1/2},&\quad \mbox{ for } r<\alpha,\\
0,&\quad\mbox{ for } r\ge \alpha, 
\end{array}\right.
\end{eqnarray}
where $r=\sqrt{x^2+y^2}$, $\sigma_0$ is the surface density at the disk center, $\alpha$ is a prescribed constant, and $n$ represents the order of the disk. 
The corresponding potential associated with a disk, $\sigma_{D_n}$, is described in Appendix A by a set of recursive formulae. 
Numerical results obtained from the proposed method are compared with the analytic solutions associated with 
$\sigma_{D_5}$ disks and are presented below. 

It is known that the complexity of the calculation of a convolution is nearly
linear $O(M\log M)$, where $M$ is the length of vectors.
The numerical complexity of the proposed method
is linear $O(N)$ for the calculation of force kernel ${\cal K}^{x,\cdot,\ell,\hat\ell}$,
where $N$ is the total number of cells.
It follows that the total complexity is nearly linear $O(N\log N)$.
Thus, we focus on exploring the order of accuracy associated with the proposed method. 

Let us define the $p$-norm of a function as
\begin{eqnarray*}
\|f\|=\left(\int_\Omega |f(x)|^p\right)^{1/p},\quad \mbox{ if }{ p\ge 1}
\end{eqnarray*}
and 
\begin{eqnarray*}
\|f\| = \mbox{ess max}_\Omega |f(x)|,\quad \mbox{ if } p=\infty.
\end{eqnarray*}
The errors between the analytic and numerical solution for various resolutions using different norms ($L^1$, $L^2$ and $L^{\infty}$) demonstrate different senses of convergence. $L^1$ norm measures the variation of errors. $L^2$ error norm is often associated with the energy involved in the errors. For example, the integral of the square of the electric field $\int E(x)^2 {\rm d }x$, i.e., the square of the $L^2$ norm, is the energy stored in the electric field. $L^{\infty}$ norm measures the errors in a pointwise sense which is a strong sense of convergence. We apply those definitions of norm to the following examples and show that the proposed method is of second order accuracy. 

\hhw{The common properties of} \rt{the} \hhw{ examples discussed below are summarized as follows: (1) The disk model $\sigma_{D_5}$ 
with a specific $\alpha$, which describes the size of the disk, is adopted to construct either a monopole (examples 1, 2) or a 
dipole field (examples 3, 4) to demonstrate the order of accuracy of our algorithm. (2) The one-side cell size of those cells 
in $\Omega^{\ell}_{0}$ is one-half of that in $\Omega^{\ell - 1}_{0}$. That is, the cell size ratio used between levels 
$\ell - 1$ and $\ell$ is taken to be 2. (3) The cell size of $\Omega^0_{0}$ is described by $(\Delta x) = (\Delta y)=(1/2)^{k-1}$, 
with $k=5,\ldots,10$, corresponding to the one-side cell number $N=32, \ldots, 1024$. We note that the number of \hhwang{cells} $N$ is 
applied to the root level, i.e., $\ell = 0$. The cells of all levels of refinement are half sized as $N$ is doubled. (4) 
Figures~1 to 4 corresponds to Examples 1 to 4 described below. In each figure, the mesh configuration is schematically shown in 
the top-left panel, the top-right panel shows the contours of the surface density, and the contours of absolute errors for 
$x$-force and $y$-force between analytic and numerical solutions are shown in the bottom-left and bottom-right panels in the 
common logarithmic scale, respectively.  (5) Tables~1 to 4 show the corresponding $L^1$, $L^2$, and $L^\infty$ errors between 
analytic and numerical solutions for $k=5,\ldots,10$. }

\begin{enumerate}
\item[{\sl Example 1.}] 
We demonstrate the order of accuracy using a $\sigma_{D_5}$ disk with $\alpha=0.85$. The disk is centered at the origin of coordinates and covered with two levels of nested grids. The domains of \hhw{the grids are the following:} $\Omega^1_0=[-1/2,1/2]\times[-1/2,1/2]$, and $\Omega^0_0=[-1,1]\times[-1,1] \cap (\Omega^1_0)^c$.  The \rt{corresponding} errors between analytic and numerical 
solutions for $x$-force and $r$-force are detailed in Table \ref{tblExample1}. It shows that the proposed method is almost second 
order accuracy for each norm in this example.

\item[{\sl Example 2.}] In this case, we show the nested grid calculation using three levels with $\ell=0,1,2$.  The surface 
density adopted is \rt{identical to} that used in {\sl Example 1}. The domains of \hhw{the grids are the following:} 
$\Omega^2_0=[-1/4,1/4]\times[-1/4,1/4]$, $\Omega^1_0=[-1/2,1/2]\times[-1/2,1/2] \cap (\Omega^2_0)^c$ and $\Omega^0_0=[-1,1]\times[-1,1]\cap(\Omega^2_0\cup\Omega^1_0)^c$. 
The corresponding errors between analytic and numerical solutions for $x$-force and $r$-force are detailed in Table \ref{tblExample2}. It shows that the proposed method is also almost second order accuracy for each norm in this example.  

\item[{\sl Example 3.}] The non-axisymmetric case consists of two $\sigma_{D_5}$ disks
with $\alpha=\sqrt{2}/4$ \hhw{and} is demonstrated for a two-level nested grid simulation.
The centers of two disks are located at $(-1/2,0)$ and $(1/2,0)$. 
The domains of \hhw{the grids are the following:} $\Omega^1_0=[-0.75,-0.25]\times [-0.25,0.25]$
, $\Omega^1_1=[0.25,0.75]\times [ -0.25,0.25]$ and $\Omega^0_0=([-1,1]\times [-1,1])\cap (\Omega^1_0\cup \Omega^1_1)^c$.
The corresponding errors between analytic and numerical solutions
for $x$-force and $y$-force are detailed in Table \ref{tblExample3}. 
It shows that the order of accuracy is $2$ for $L^1$ and $L^2$ norms, while $1.9$ for $L^\infty$ norm.

\item[{\sl Example 4.}] 
The non-axisymmetric distribution of surface density is the same as that used
in Example 3. In this case, the disks are covered with three-level nested grids. 
The domains of \hhw{the grids are the following:} For level $\ell=2$, $\Omega^2_0=[0.25, 0.75]\times [-0.25,0.25]$
and $\Omega^2_1=[-0.75,-0.25]\times [-0.25,0.25]$; for level $\ell=1$, $\Omega^1_0=[-0.9, -0.1]\times [-0.4,0.4]\cap (\Omega^2_1)^c$ and $\Omega^1_0=[0.1, 0.9]\times [-0.4,0.4]\cap (\Omega^2_0)^c$; and for $\ell=0$, $\Omega^0_0=[-1,1]\times[-1,1]\cap(\Omega^2_0\cup\Omega^2_1\cup\Omega^1_0\cup\Omega^1_1)^c$. 
The corresponding errors between analytic and numerical solutions
for $x$-force and $y$-force are detailed in Table \ref{tblExample4}. It shows that the order of accuracy is $2$ for $L^1$ and $L^2$ norms, while $1.9$ for $L^\infty$ norm.
 
\end{enumerate}

\section{Conclusions and discussions}

We have extended the method developed by \citet{Yen2012} for directly calculating the 
self-gravity force \hhw{in an infinitesimally thin disk, i.e., in two dimensions, induced} from a surface density on a nested grid configuration. 
It is worth noting that the direct approach is to represent the forces in a convolution form 
on uniform/nested grids. 
Therefore, the fast Fourier transform is employed only for speeding up the computational time or reducing the numerical complexity. 
The method also has been demonstrated to be of second order accuracy by an analytic potential-density pairs.

In practice, the implementation of the proposed method for (\ref{eqnCalK}) can be modified as
\begin{eqnarray}
\label{eqnCalKepsilon}
{\cal K}^{x,0,\ell,\hat\ell} (x,y)
=-|\mbox{sgn}(x)|\ln(y+\sqrt{x^2+y^2}+\epsilon) 
-(1-|\mbox{sgn}(x)|)\mbox{sgn}(y) \ln (|y|+\epsilon),
\end{eqnarray}
where $\epsilon$ is a positive number less than the round off errors. 
We note that the number $\epsilon$ is only introduced in numerical calculations to avoid 
numerical Not-a-Number (NaN) errors. 
The value of $\epsilon$ is of the order of the machine error and should not be taken as a softening length.  

\begin{figure}
\begin{center}
\includegraphics[width=.45\textwidth]{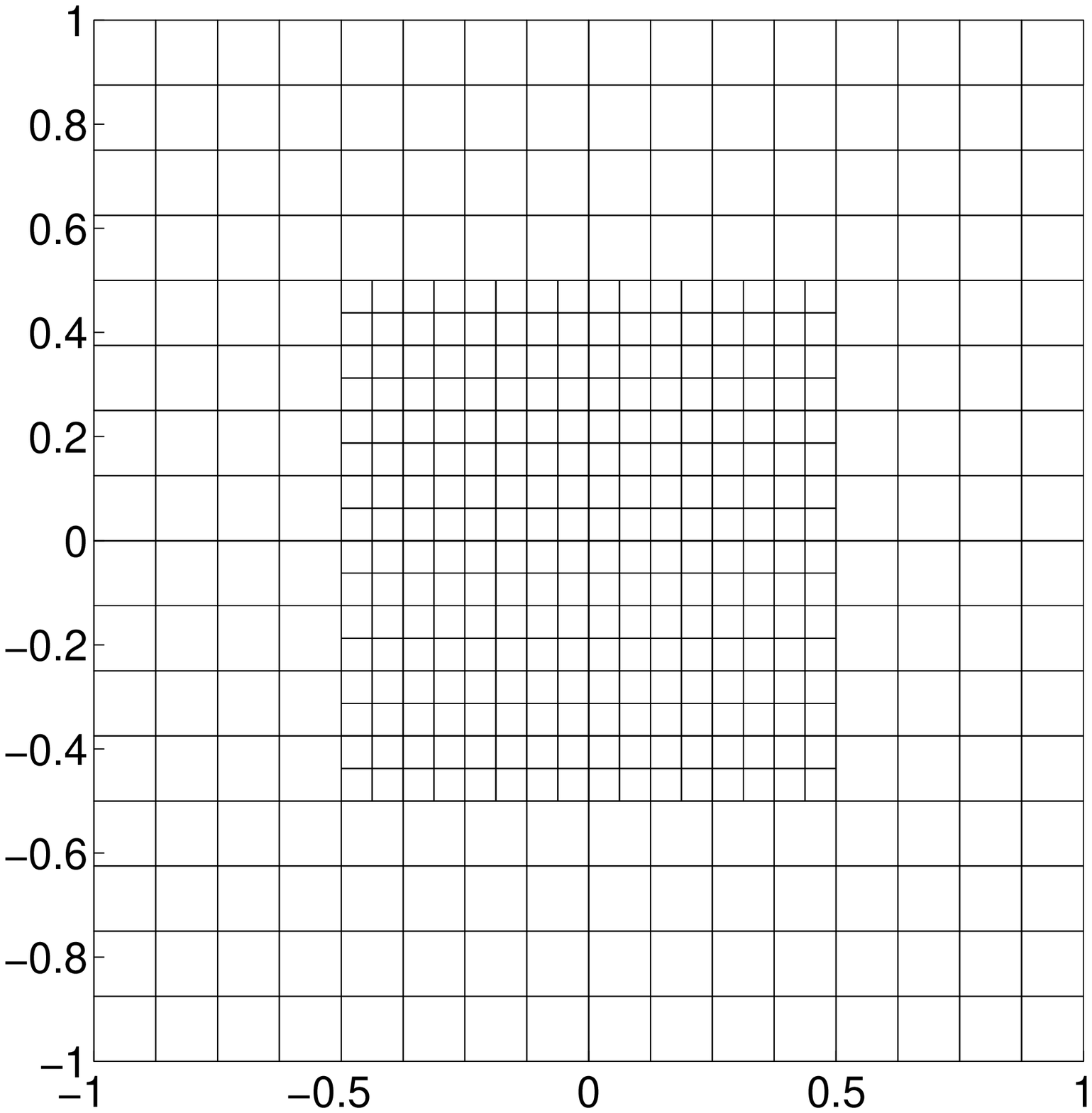}\quad
\includegraphics[width=.45\textwidth]{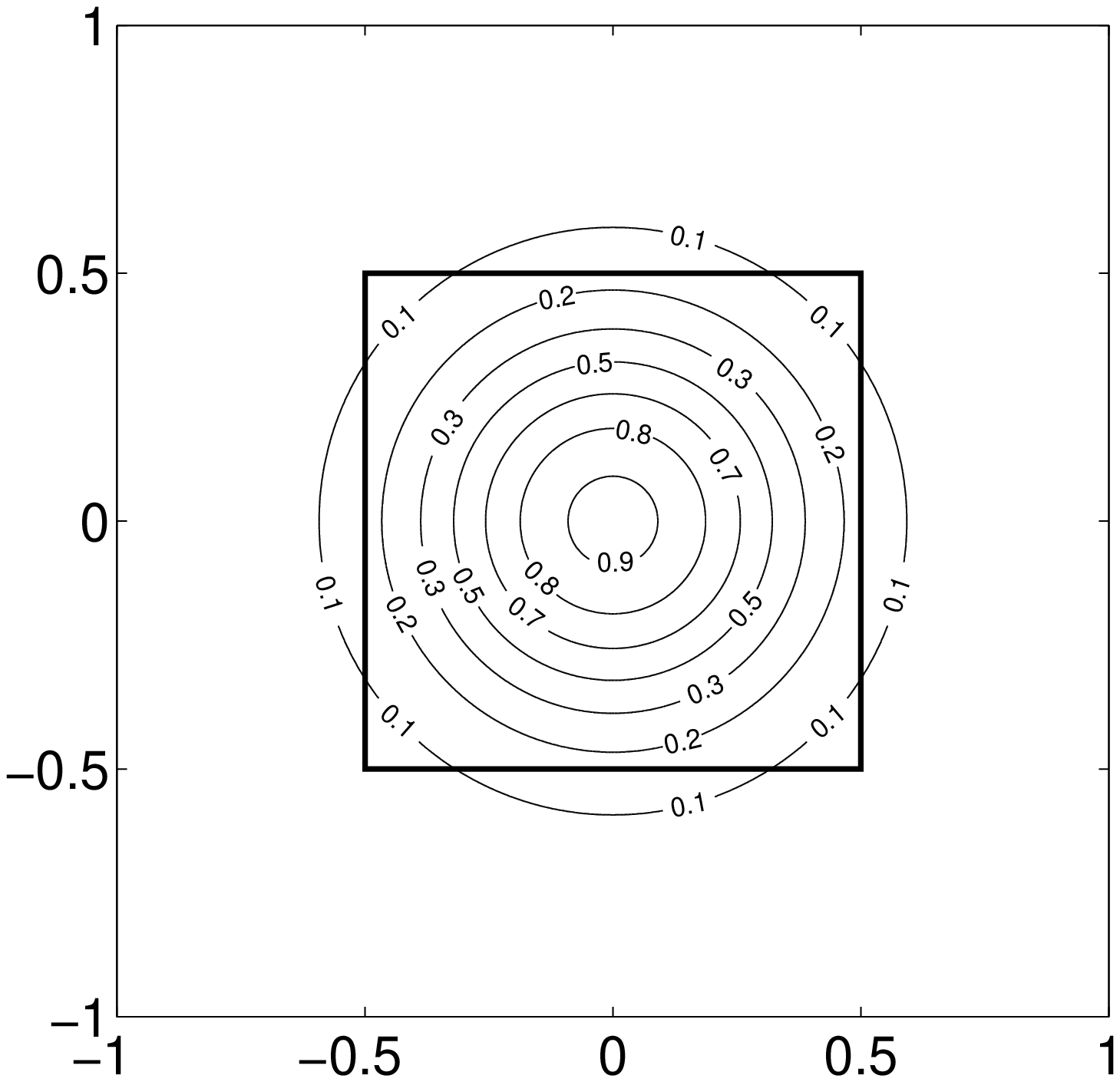}\\
\includegraphics[width=.45\textwidth]{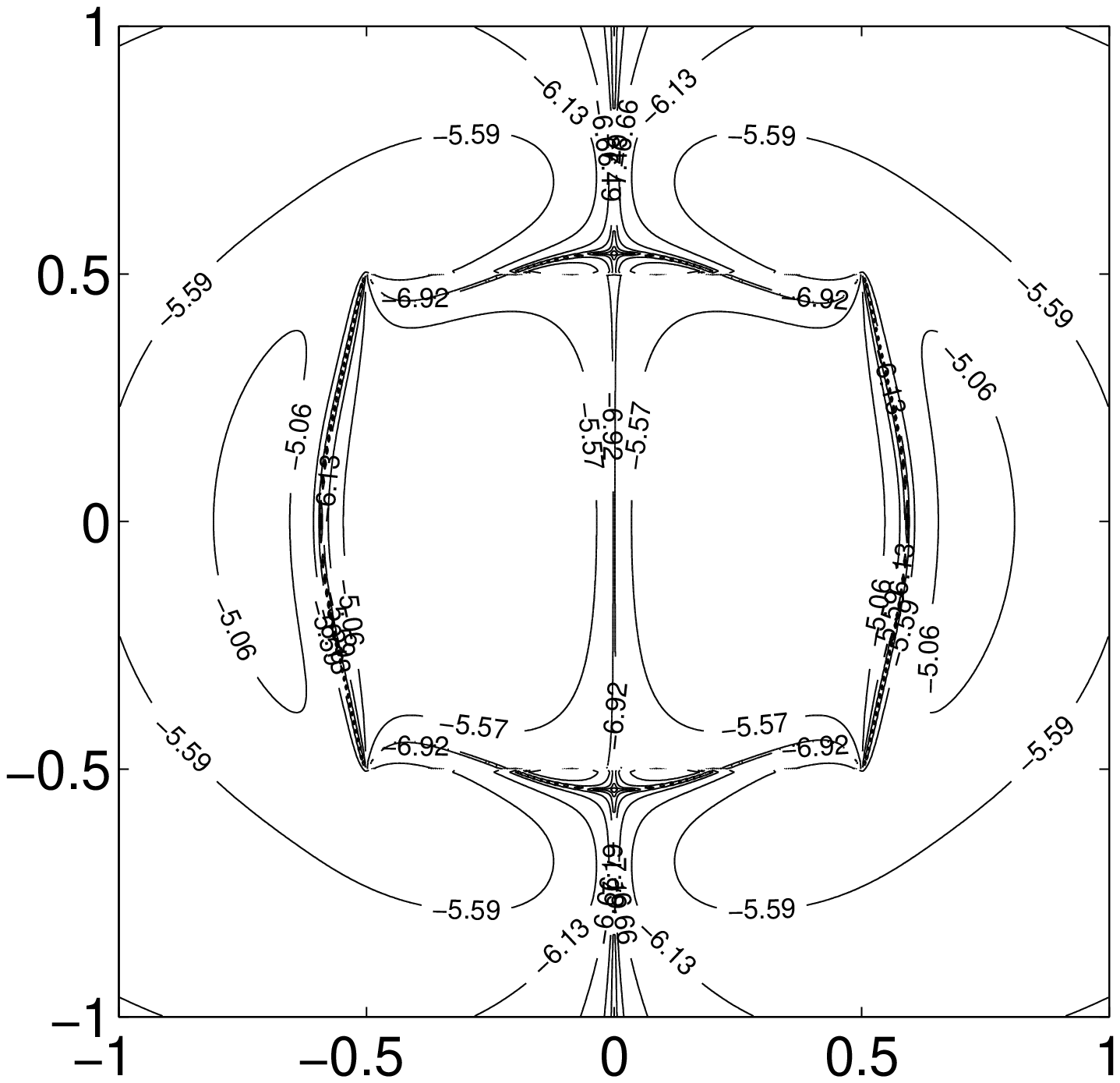}\quad
\includegraphics[width=.45\textwidth]{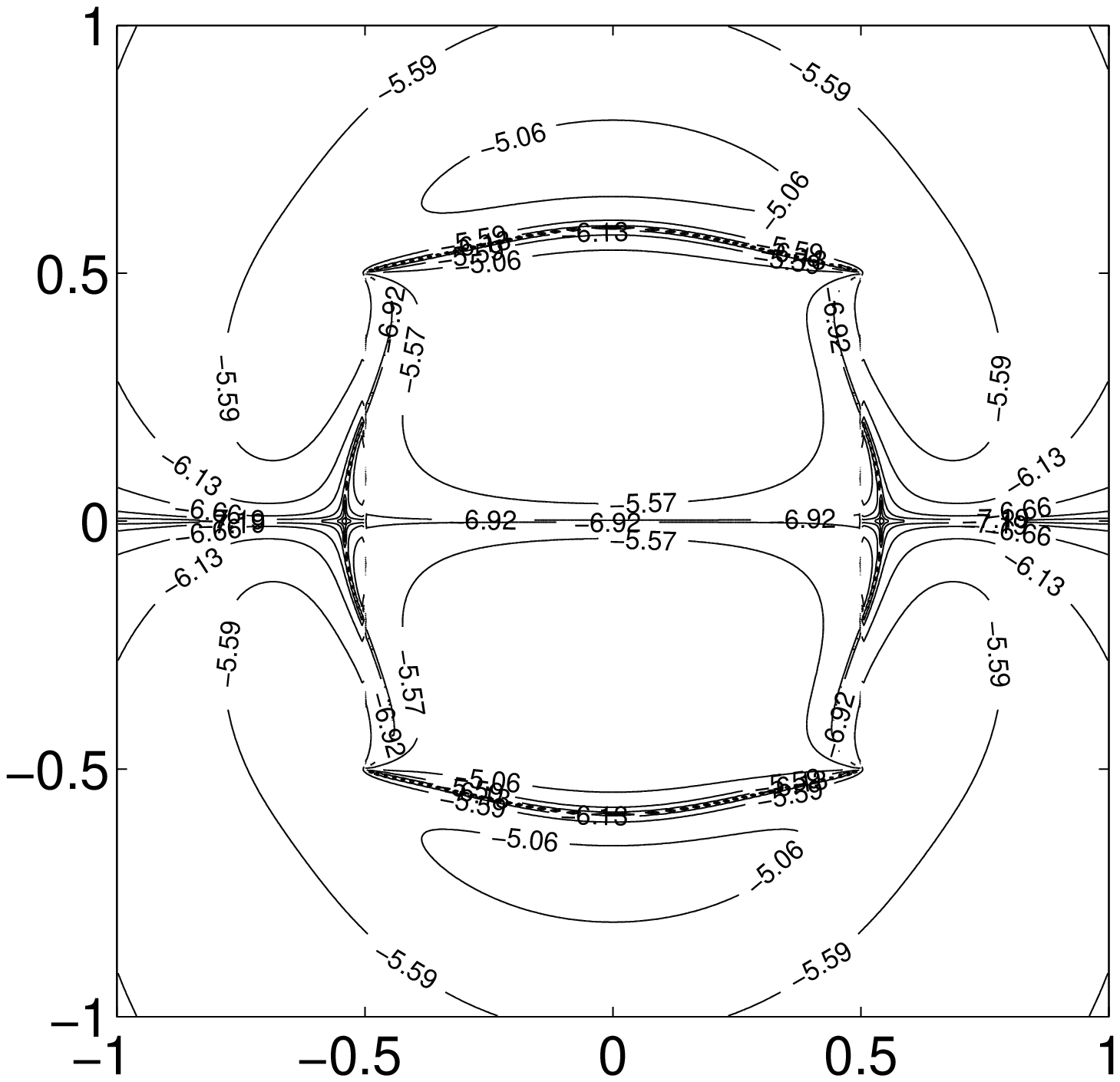}
\end{center}
\caption{The numerical simulations of a $\sigma_{D_5}$ disk for $N=512$
based on two grid levels. The relation between the coarser and finer grids is sketched in top-left panel.
The contour plot in the top-right panel is the surface density. 
The square with thick line shows the boundary of the refined grid. 
The corresponding errors between the analytic 
and numerical solutions are shown for $x$-force (bottom-left) and 
$y$-force (bottom-right). The values in the contour plots in the bottom row are the absolute 
errors in the common logarithmic scale.}
\label{Fig1}
\end{figure}

\begin{figure}
\begin{center}
\includegraphics[width=.45\textwidth]{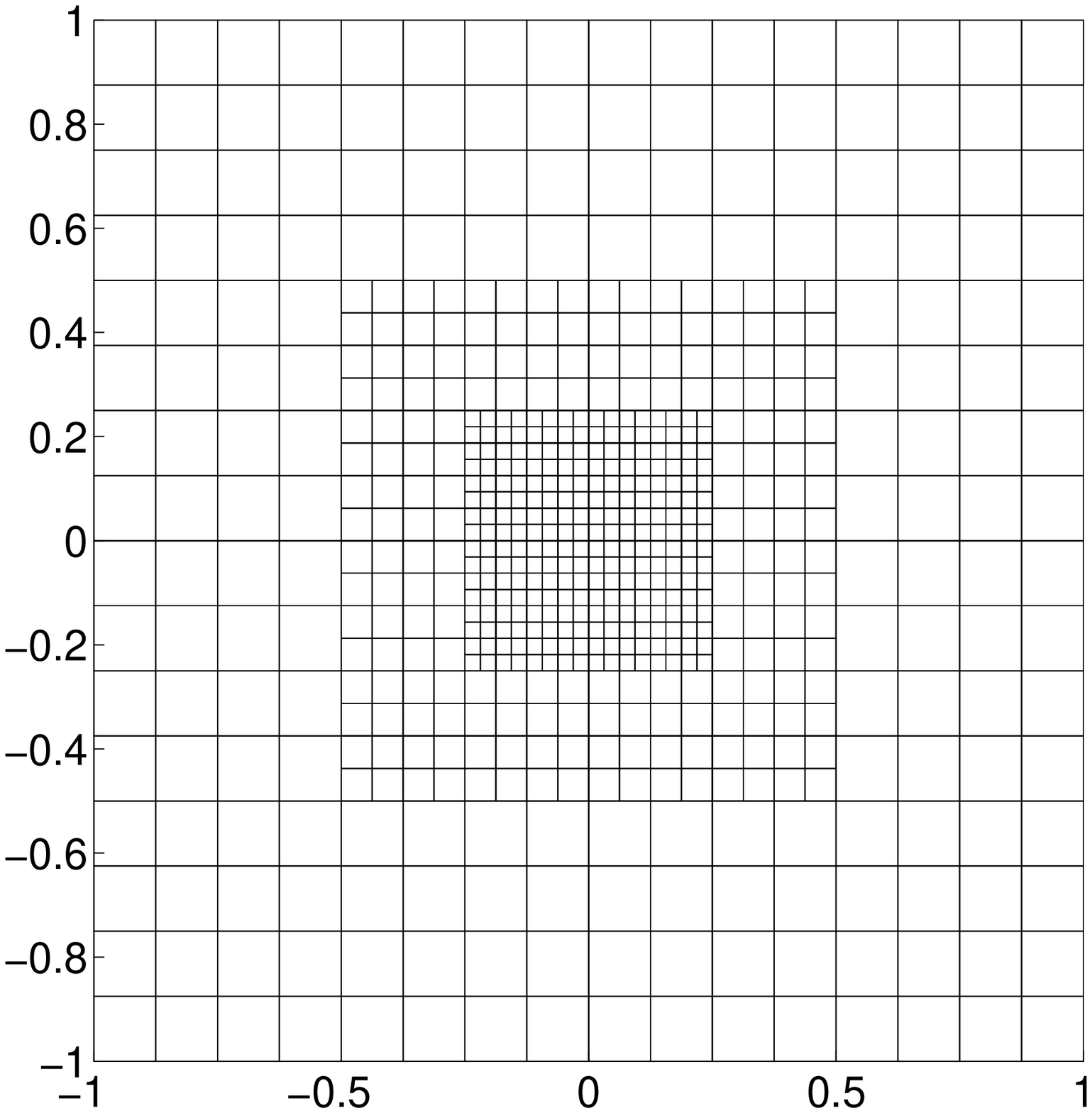}\quad
\includegraphics[width=.45\textwidth]{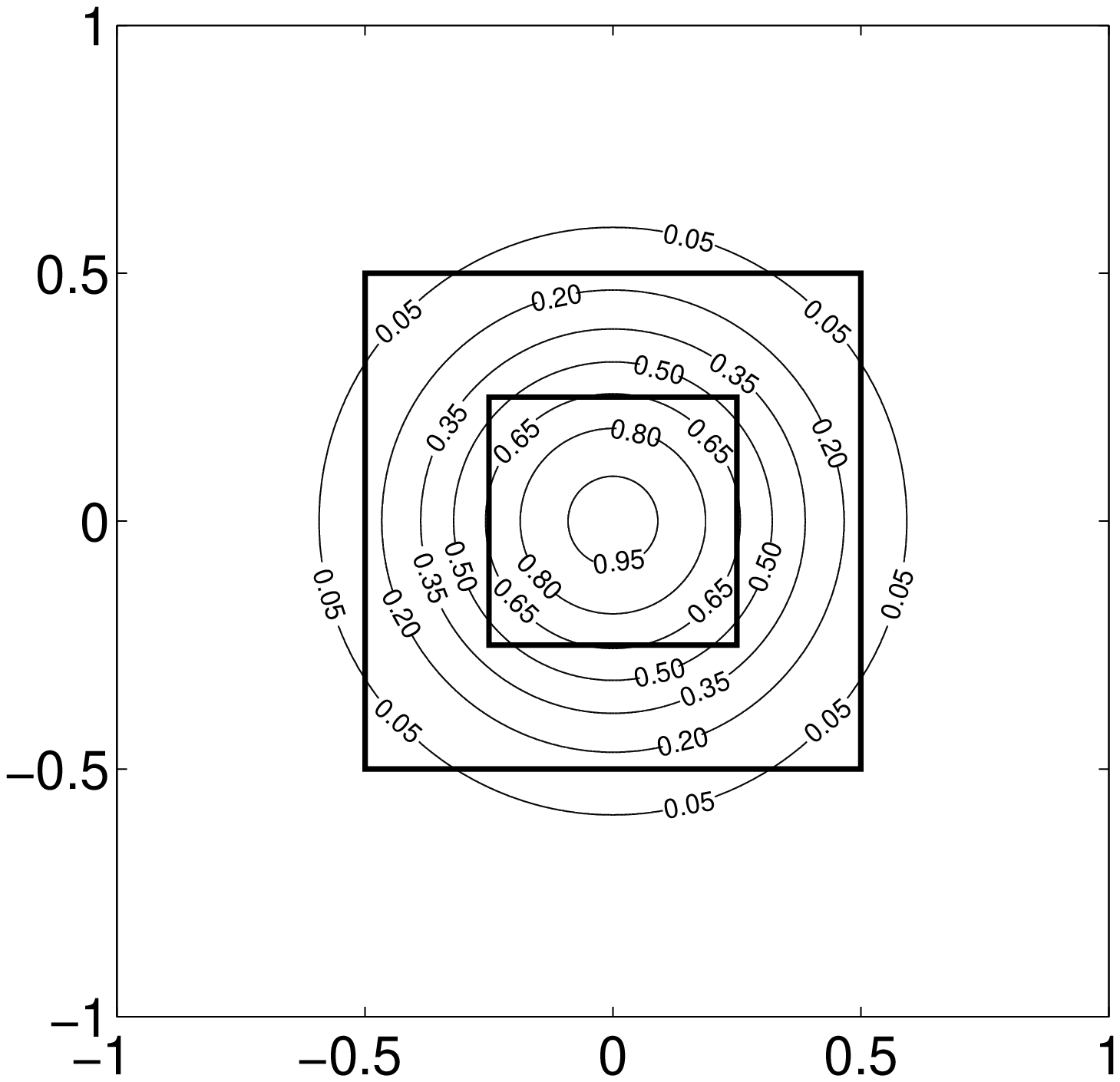}\\
\includegraphics[width=.45\textwidth]{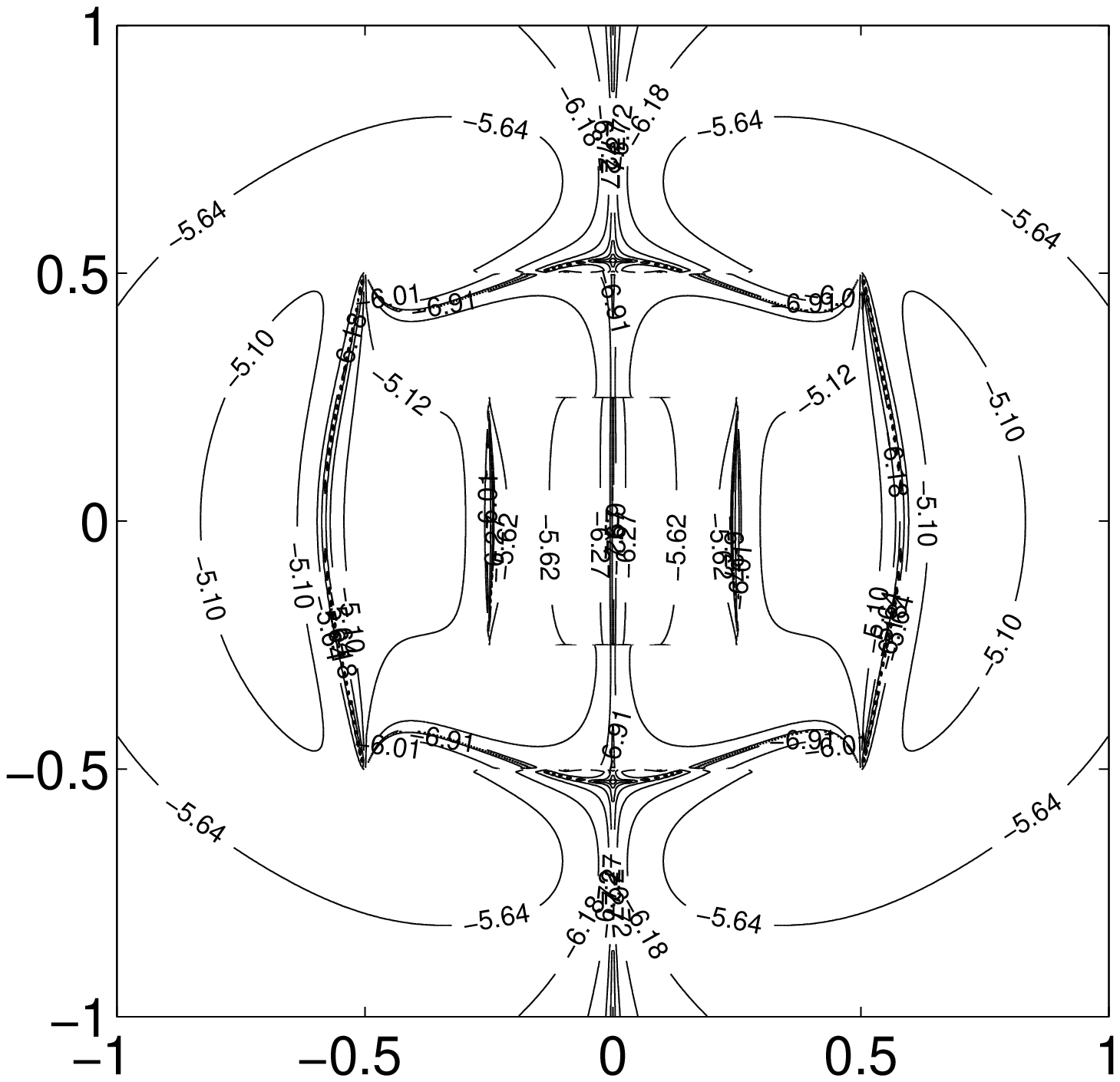}\quad
\includegraphics[width=.45\textwidth]{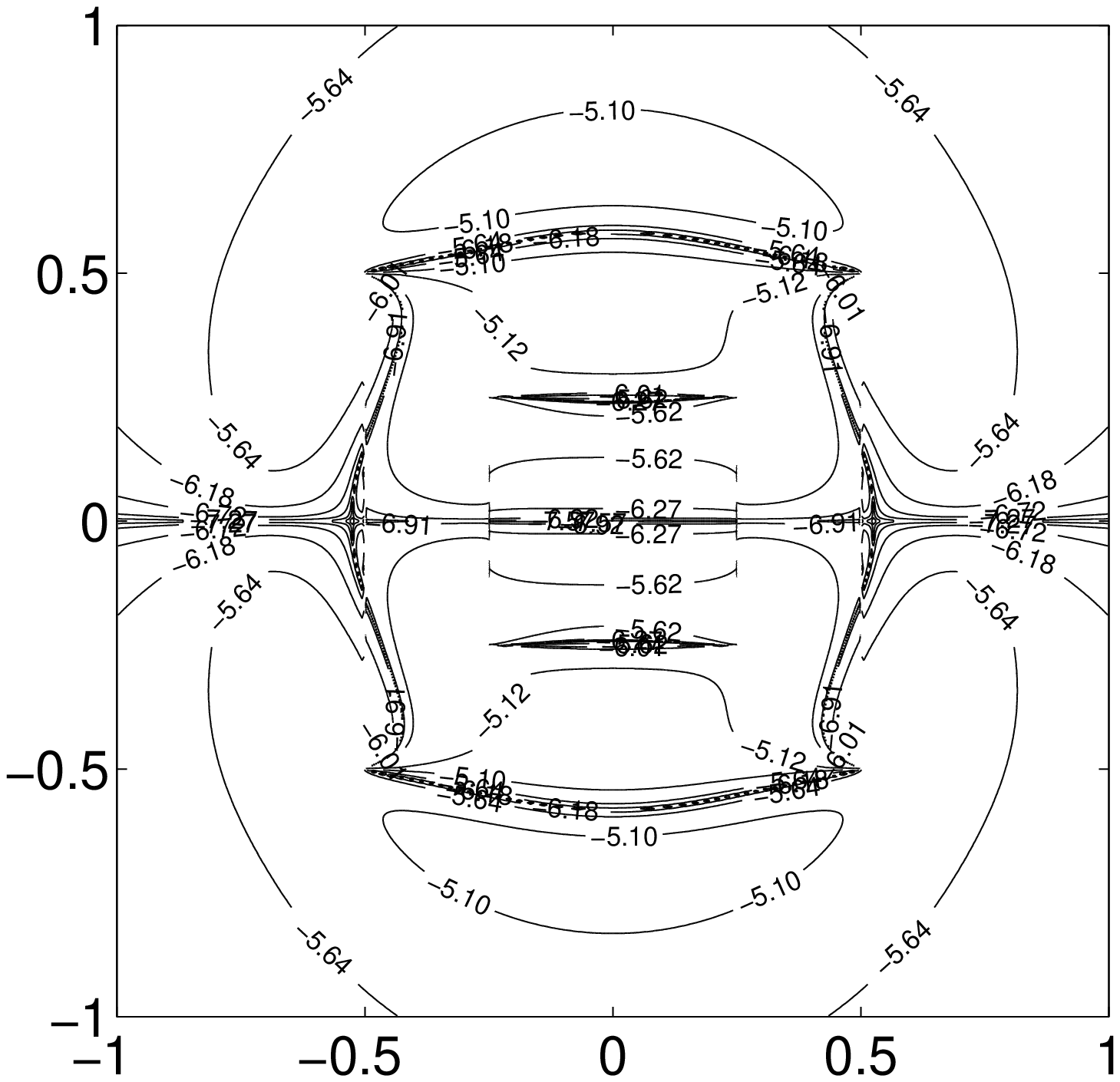}
\end{center}
\caption{The numerical simulations of a $\sigma_{D_5}$ disk for $N=512$
based on three grid levels, $\ell=0,1,2$. The relation among the grid levels is sketched in top-left panel.
The contour plot in the top-right panel is the surface density. 
The squares with thick line show the boundaries of the refined grids.
The corresponding errors between the analytic 
and numerical solutions are shown for $x$-force (bottom-left) and 
$y$-force (bottom-right). The values in the contour plots in the bottom row are the absolute 
errors in the common logarithmic scale.}
\label{Fig2}
\end{figure}

\begin{figure}
\begin{center}
\includegraphics[width=.45\textwidth]{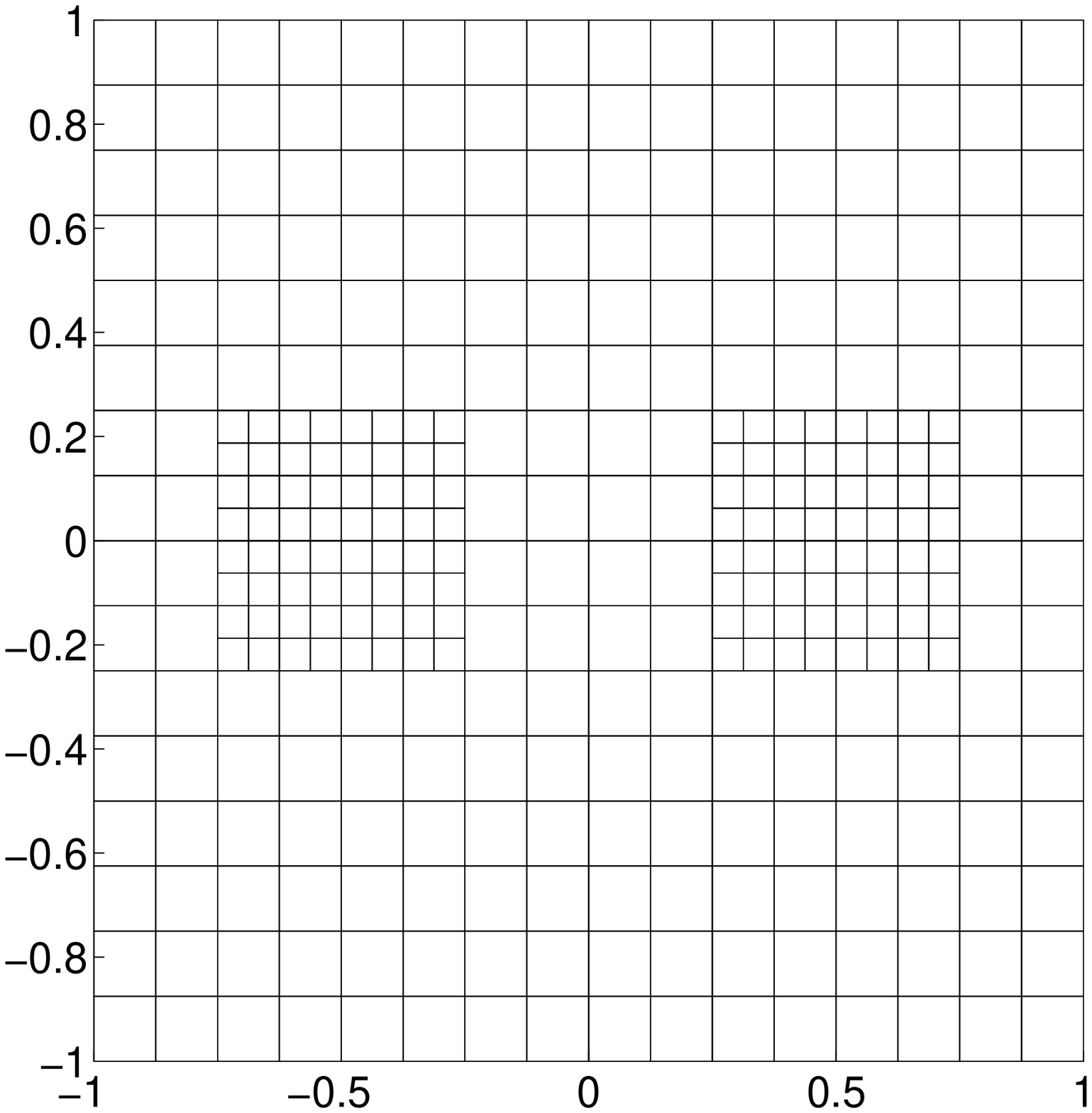}\quad
\includegraphics[width=.45\textwidth]{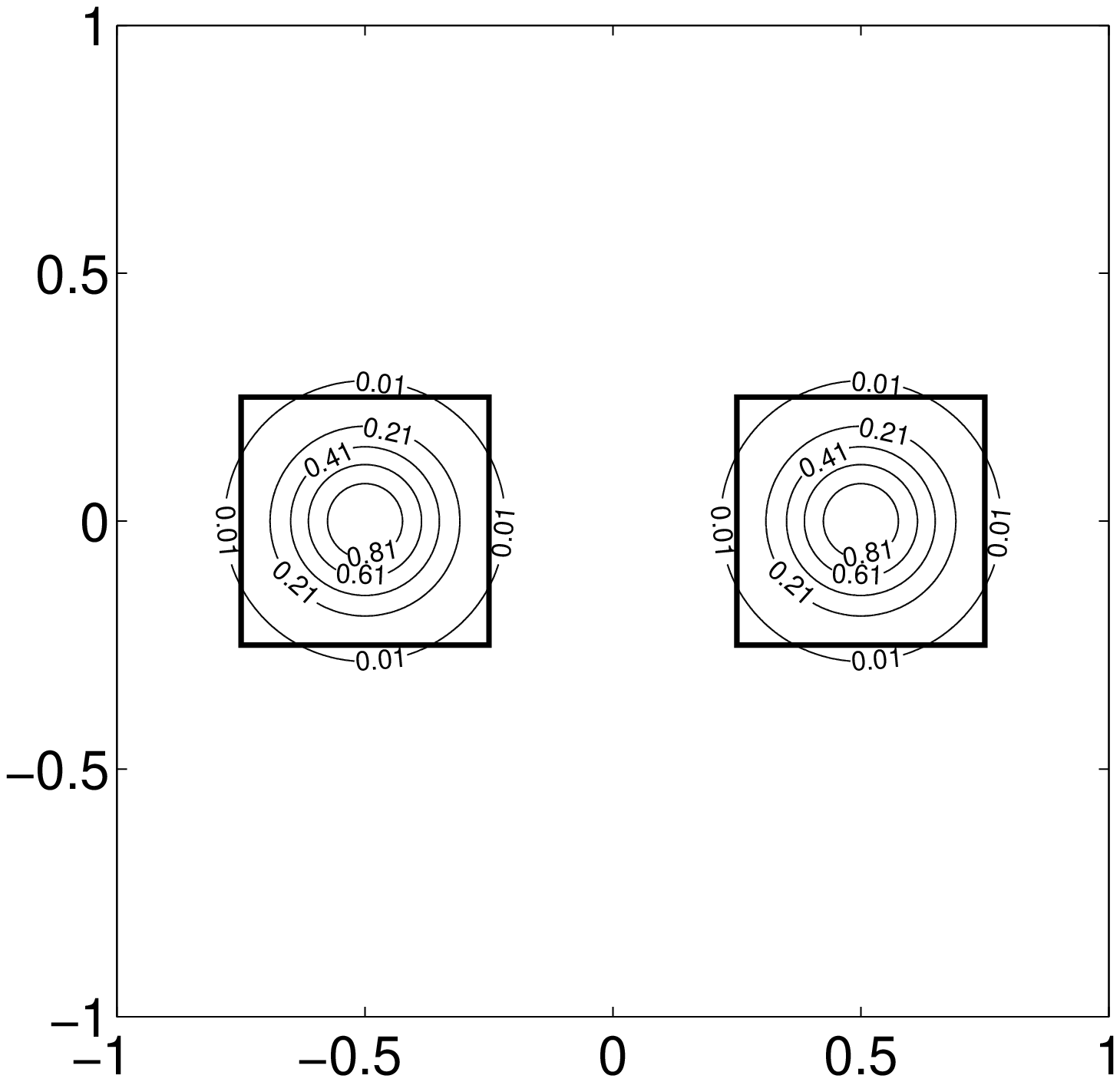}\\
\includegraphics[width=.45\textwidth]{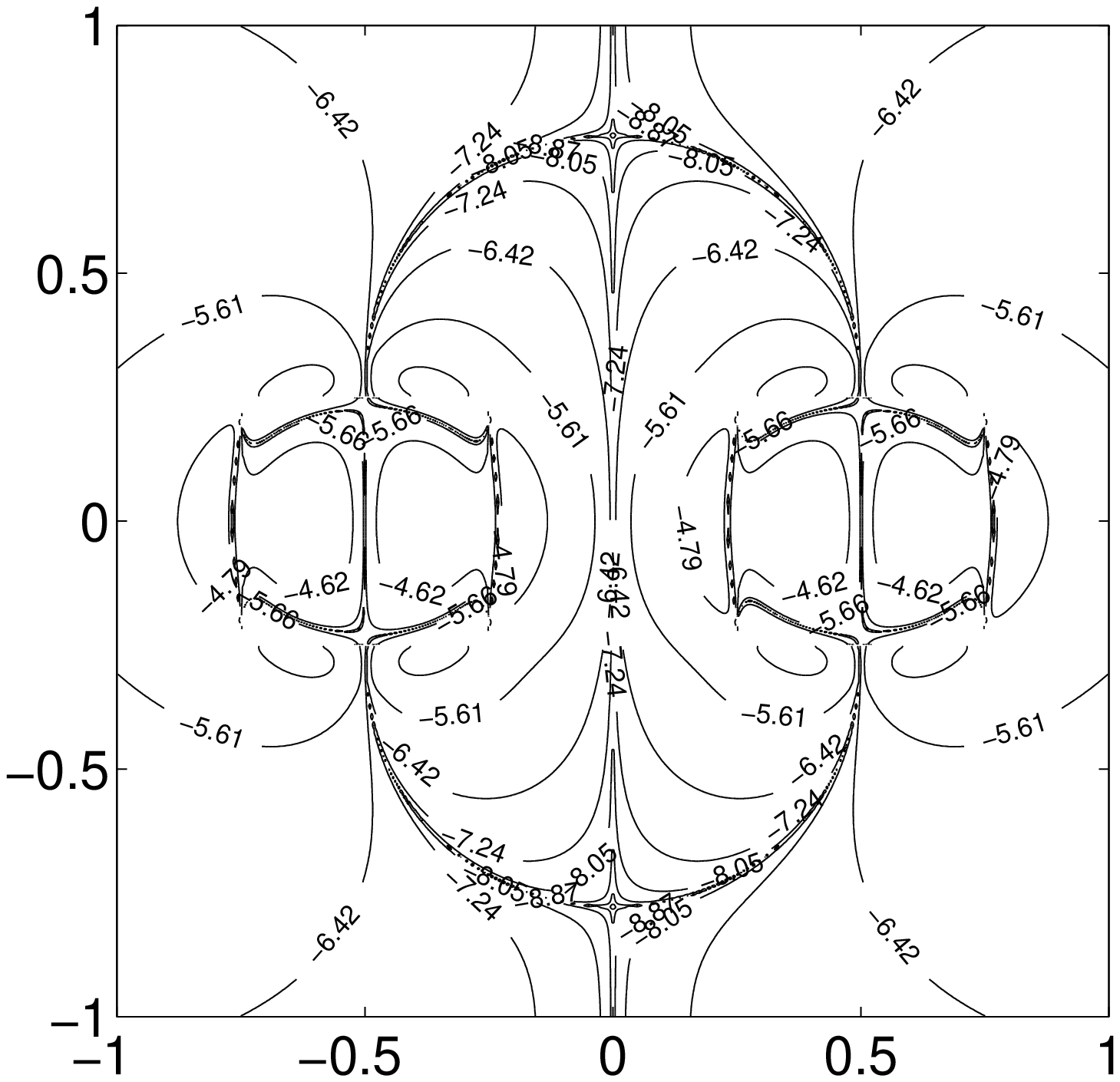}\quad
\includegraphics[width=.45\textwidth]{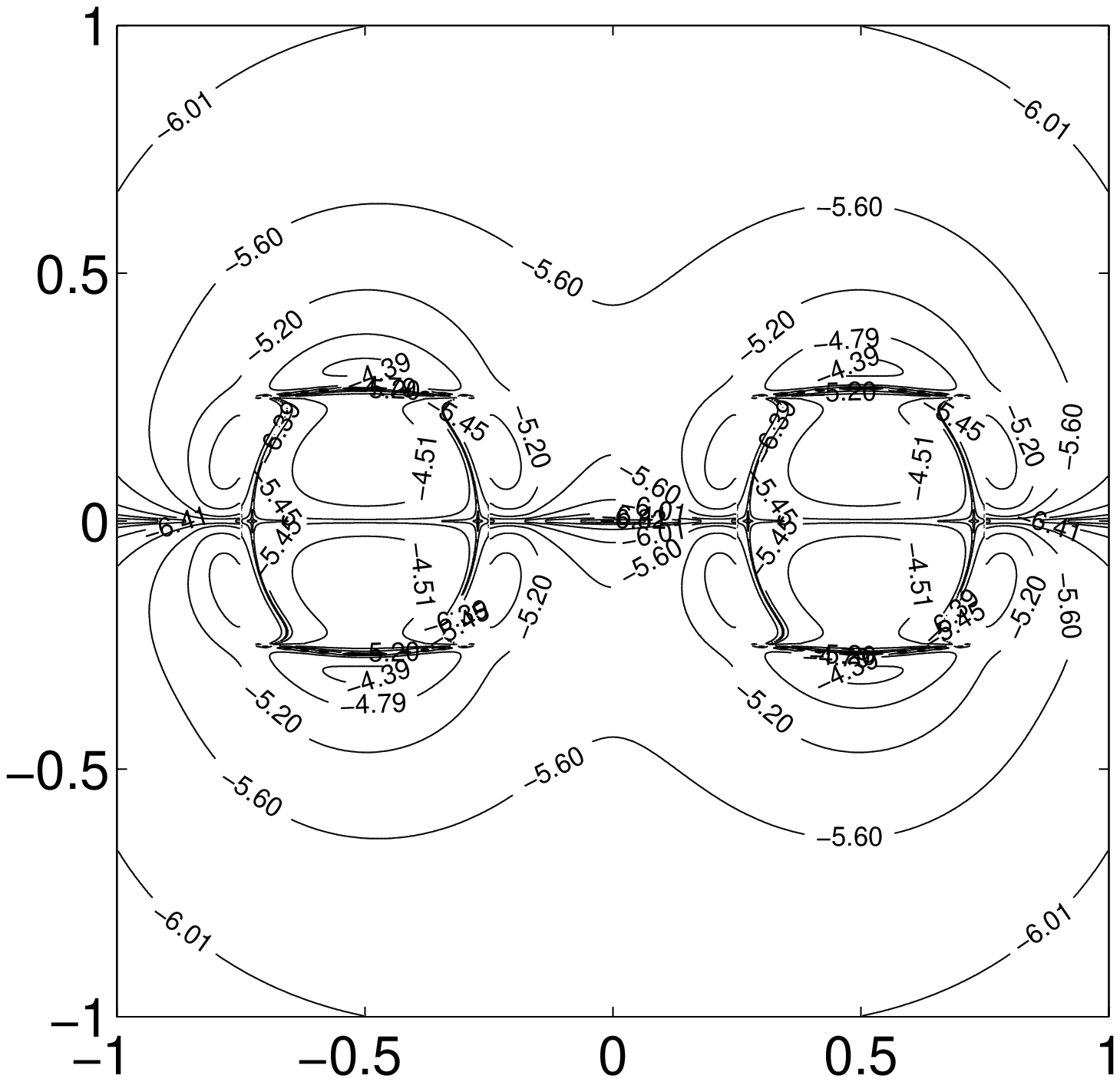}
\end{center}
\caption{The numerical simulations of a non-axisymmetric case 
which consists of two $\sigma_{D_5}$ disks with $\alpha=\sqrt{2}/4$ 
and centers at $(-1/2,0)$ and $(1/2,0)$ for $N=512$ based on two grid levels. 
The relation between the coarser and finer grids is sketched in top-left panel.
The contour plot in the top-right panel is the surface density.
The squares with thick line show the boundaries of the refined grids. 
The corresponding errors between the analytic 
and numerical solutions are shown for $x$-force (bottom-left) and 
$y$-force (bottom-right). The values in the contour plots in the bottom row are the absolute 
errors in the common logarithmic scale.}
\label{Fig3}
\end{figure}

\begin{figure}
\begin{center}
\includegraphics[width=.45\textwidth]{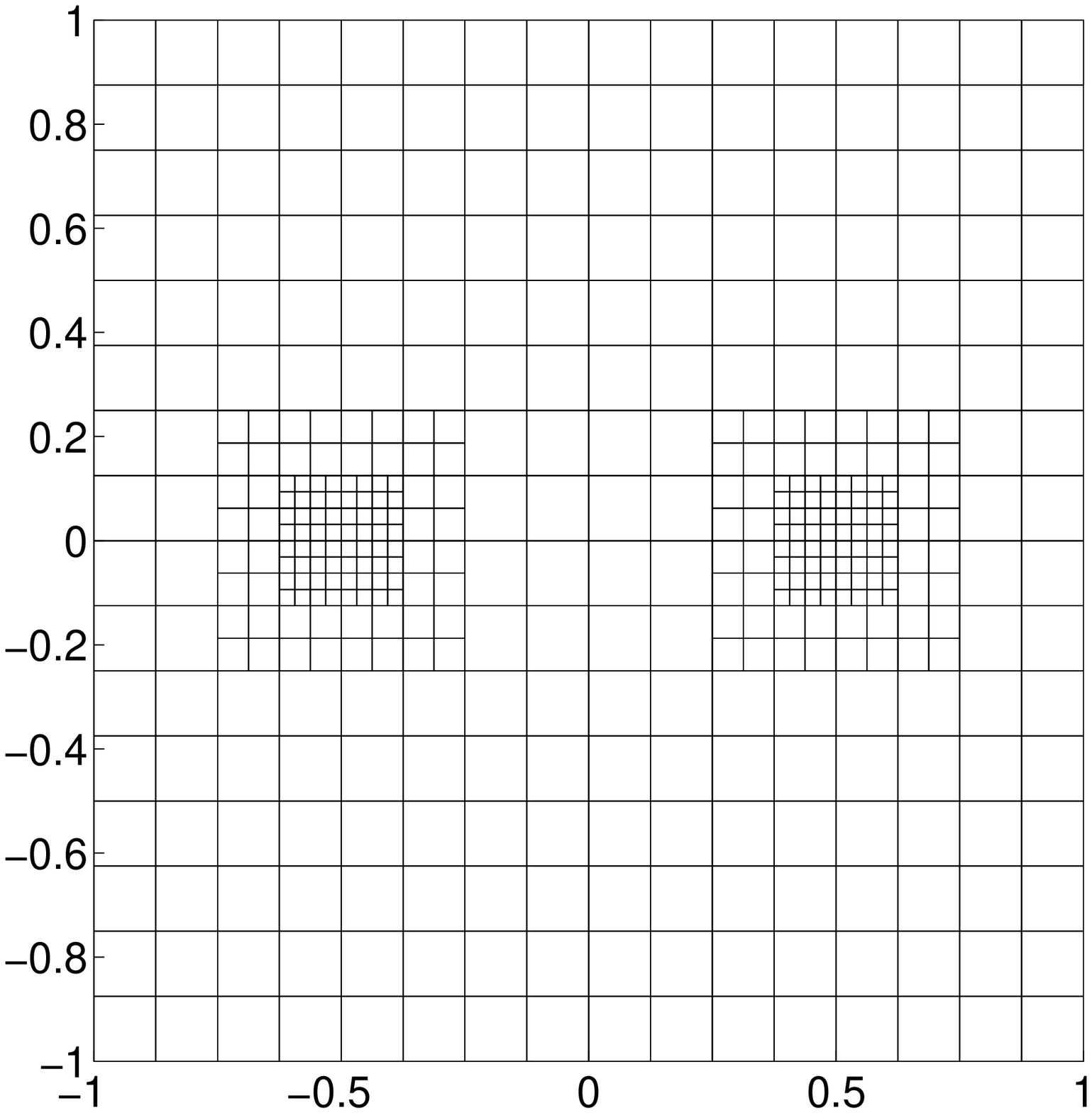}\quad
\includegraphics[width=.45\textwidth]{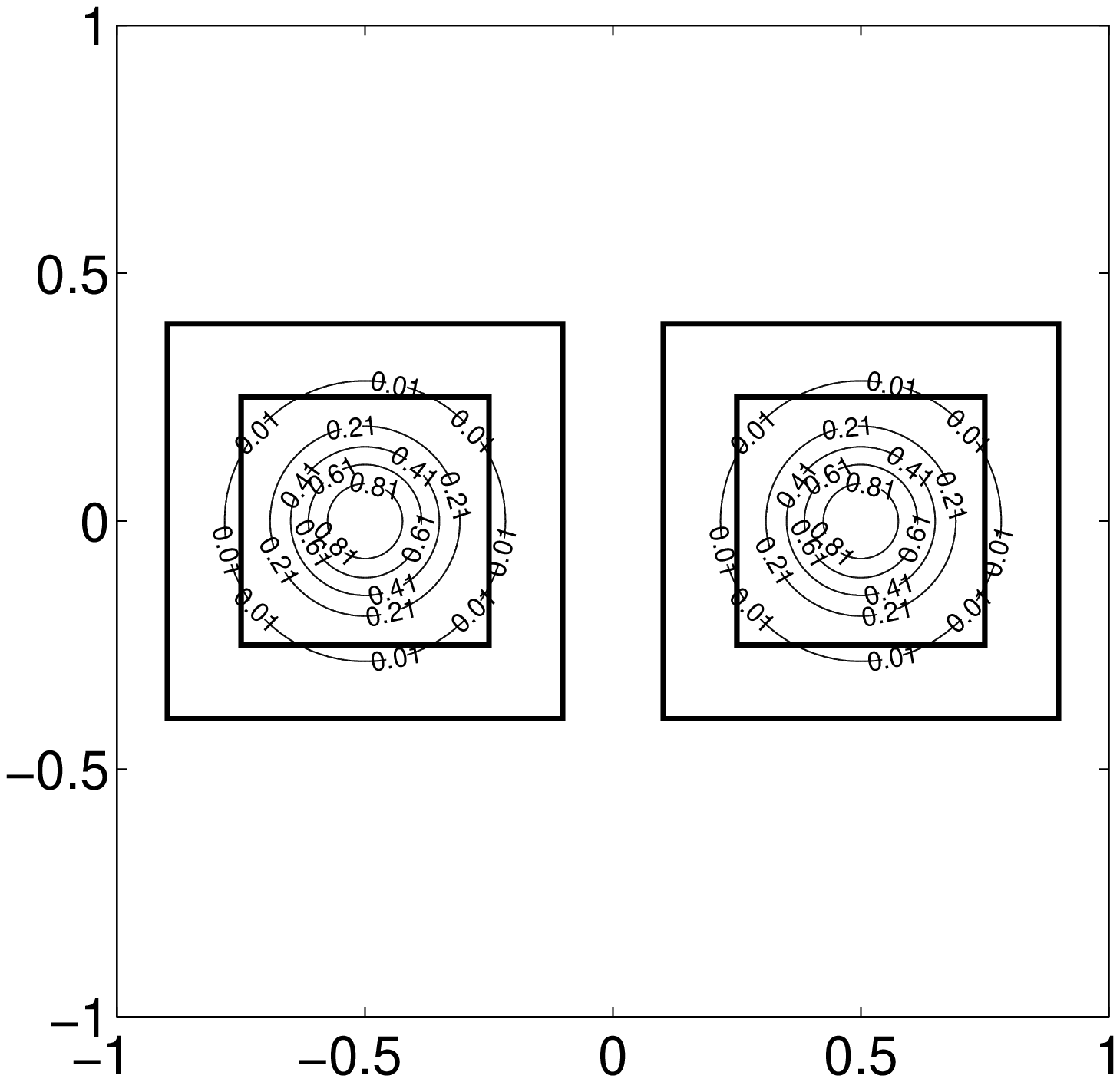}\\
\includegraphics[width=.45\textwidth]{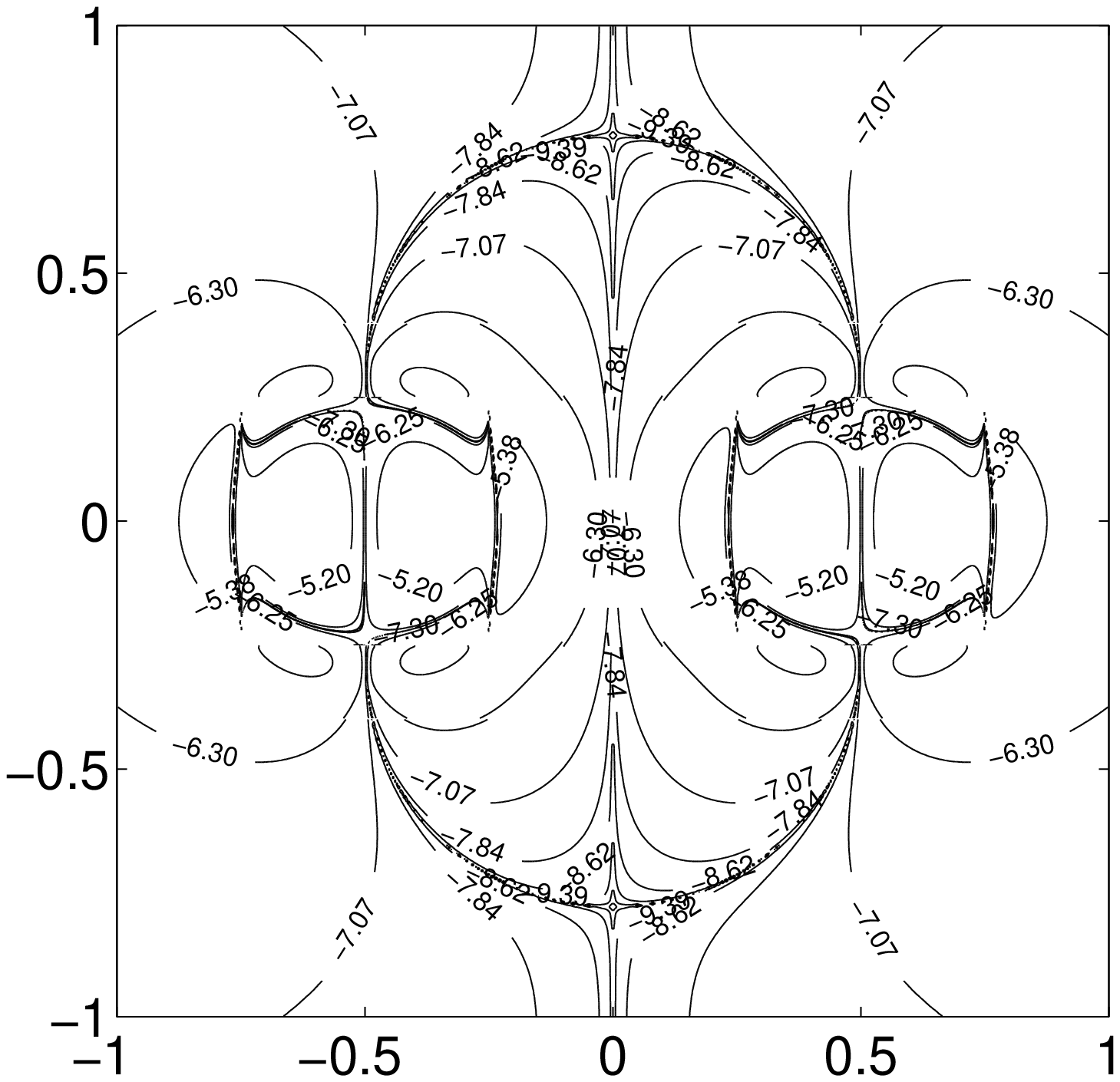}\quad
\includegraphics[width=.45\textwidth]{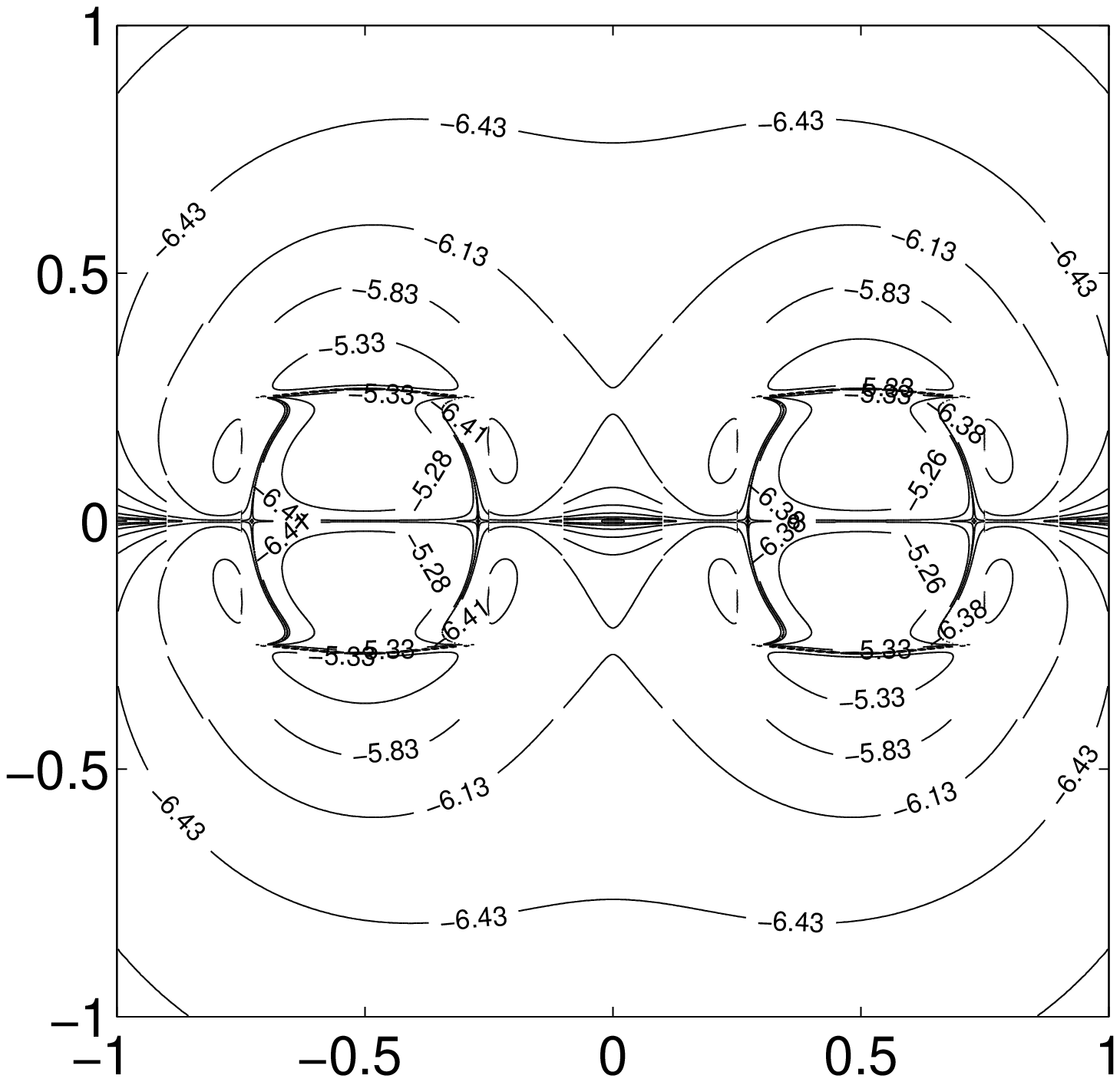}
\end{center}
\caption{The numerical simulations of a non-axisymmetric case 
which consists of two $\sigma_{D_5}$ disk with $\alpha=\sqrt{2}/4$ 
and centers at $(-1/2,0)$ and $(1/2,0)$ for $N=512$
based on three grid levels, $\ell=0,1,2$.
The relation among the grid levels is sketched in top-left panel.
The contour plot in the top-right panel is the surface density. 
The squares with thick line show the boundaries of the refined grids.
The corresponding errors between the analytic 
and numerical solutions are shown for $x$-force (bottom-left) and 
$y$-force (bottom-right). The values in the contour plots in the bottom row are the absolute 
errors in the common logarithmic scale.}
\label{Fig4}
\end{figure}

As shown in the bottom row of Figures~\ref{Fig1} to \ref{Fig4}, the major errors are concentrated in the 
immediate vicinity of the interfaces where the spatial resolution undergoes a transition. This degrades 
the order of accuracy to roughly 1.9 in terms of the $L^{\infty}$ norm. This actually demonstrates the 
advantage of using the integral form (\ref{eqnPhixy0G}), since the density distribution is treated smoothly when crossing the interface. Strong spurious forces at the boundaries would be visible if the mass enclosed within a cell is treated as a particle located at the cell center.  

\hhw{It is expected that the inclusion of mesh refinement will improve the numerical accuracy in comparison with the case without refinement, since more details are taken into account in the refined grids. To} \rt{determine the} \hhw{benefit from mesh refinement, 
the errors within the domain defined by $\Omega^1_0$ in {\sl Example 1}} \rt{are} \hhw{analyzed for $N=512$ with and without 
refinement. For this specific case, the improvement in terms of $L^1$, $L^2$ and $L^{\infty}$ are 1.57, 1.58, -0.22, respectively. 
This means that the accuracy is overall significantly improved, but slightly degraded in a pointwise sense. We do not expect the 
overall improvement can be second order since only a portion of \hhwang{the} disk is refined. The slightly negative \hhwang{value} of $L^{\infty}$ is also 
expected due to the presence of \hhwang{the} interface between levels. This work, however, shows that the errors induced by the interface 
between levels \hhwang{dissipates} at a speed of nearly second order as $N$ increases. }

\hhw{In comparison with the work of \citet{Yen2012}, a model using $\sigma_{D_2}$ disk with other parameters} \rt{identical to} 
\hhw{that of {\sl Example 1} is explored. With $\alpha=0.85$, 
the size of the disk is larger than the domain of refinement, i.e., only a part of the disk is refined. The corresponding error analyses are detailed in Table~\ref{tblD2}. Similar to the conclusions of \citet{Yen2012}, the calculation with refinement shows almost second order accuracy in terms of $L^1$ and $L^2$ norms, while converges to an order $\approx 1.5$ for $L^{\infty}$ norm. However, a detailed comprison between Table~\ref{tblD2} of this work and} \rt{Table~1} \hhw{of \citet{Yen2012} shows that the mesh refinement 
not only} \rt{improves} \hhw{the numerical accuracy, it also somewhat improves the order of convergence. The reduction in $L^{\infty}$ norm for a $\sigma_{D_2}$ disk, i.e., the error in pointwise sense, is due to the singularity appearing in the second derivative 
of} \rt{the density at the edge of the} \hhw{disk. This indicates that a $\sigma_{D_2}$ by nature is not suitable for verifying a numerical method with an accuracy higher than first order. On the other hand, this also suggests the limitation of the algorithm developed in this work.  The requirement of smoothness in \hhwang{the} density distribution reduces the numerical accuracy at density discontinuities. }

\hhw{Discontinuities in density are fairly common in astrophysical environments and numerical applications. For a specific case, it is possible to improve the order of numerical accuracy around a discontinuity if the analytical form of \hhwang{the} density is known a priori , for instance, replacing the Taylor expansion by a least square approach to minimize the error to a second order accuracy at a discontinuity. Unfortunately, in general, the density distribution is a quantity that needs to be calculated numerically and determining discontinuities is not a trivial task \hhwang{either}. Furthermore, a Poisson solver is usually coupled with a (magneto-)hydrodynamical solver, which usually reduces the order of accuracy to zero order, i.e., in terms of $L^{\infty}$, around a discontinuity, improvements \hhwang{of both hydrodynamical and Poisson solvers are} desirable in the future. }

We believe that the nested grid calculation would also be of use for applications requiring 
polar coordinates.  Based on the Green's function representation of the
potential, \citet{Chan2006}  employ a pseudo-spectral method on a scaled cosine radial grid
to achieve the high order accuracy. A modification to a uniform polar grid~\cite{Li2009}
is developed for disks with vertical structures
and the associated overall computational complexity is $O(N_r N_\phi\log N_\phi+ N_\phi N^2_r)$,
where $N_r$ is the number of cells in the radial direction and 
$N_\phi$ is the number of cells in the azimuthal direction.
The method proposed here can be generalized to polar coordinates
with second order accuracy and nearly linear complexity using the methods described in \citet{Wang2015}.

\acknowledgments
The authors would like to acknowledge the support of the Theoretical Institute for Advanced Research in Astrophysics (TIARA) based in Academia Sinica Institute of Astronomy and Astrophysics (ASIAA). C.~C.~Yen is supported by Short-term Visiting Program for Domestic Scholars, Academia Sinica, Taiwan, under the Grant 104-2-1-08-22. Thanks to Mr. Sam Tseng for assistance on the computational facilities and resources (TIARA cluster). The authors thank the referee for comments that helped to improve the clarity and presentation of this paper.

\bibliography{selfgravity_NestedGrid}
\appendix
\section*{Appendix: Recursive formula for potential-density pairs of a family of finite disks}

Potential-density pairs for a family of finite disks characterized by a surface 
density $\Sigma_n=\sigma_0 (1-R^2/\alpha^2)^{n-1/2}$ \rt{is} described.  It is a generalization of 
the study for $n=0,1,2$ in~\cite{Schulz2009}. 
The potential $\Phi_{D_n}$ corresponds to the finite disks with surface density $\Sigma_n$.   
The potential $\Phi_{D_1}$ is known as the Maclaurin potential for the Maclaurin disk 
\begin{eqnarray*}
\Sigma_{\mbox{Mac}}(R;\alpha)=\left\{
\begin{array}{ll}
\sigma_0 \sqrt{1-R^2/\alpha^2}& \mbox{ for } R<\alpha,\\
0 &\mbox{ for } R>\alpha.
\end{array}
\right.
\end{eqnarray*}
It is represented  for $z=0$ as 
\begin{eqnarray*}
\Phi_{D_1}(R,0;\alpha) &=& \Phi_{\mbox{Mac}}= -\frac{\pi G\sigma_0}{2\alpha} \left[ (2\alpha^2-R^2) \sin^{-1}\frac{\alpha}{R}+\alpha \sqrt{R^2-\alpha^2}\right],\\
&=& -\frac{\pi G\sigma_0}{2\alpha}R^2 \left[ (2(\xi^2-1)+1)\sin^{-1}\xi + \xi \sqrt{1-\xi^2}\right],
\end{eqnarray*}
for $R\ge \alpha$, where $\xi=\alpha/R$ and 
\begin{eqnarray*}
\Phi_{D_1}(R,0;\alpha)=-\frac{\pi^2 \sigma_0 G} {4\alpha}(2\alpha^2-R^2),
\end{eqnarray*}
for $R\le \alpha$.

Let us first define the notation $\displaystyle H(k,m)=(-1)^{k-m}\prod^k_{j=m}\frac{2j+1}{2j+2}$.
The relation between the potentials $\Phi_{D_n}$ is 
\begin{eqnarray}
\label{eqnPhiDnp1}
\Phi_{D_{n+1}}= \frac{2n+1}{\alpha^{2n+1}} \int^\alpha_0 {\hat \alpha}^{2n}\Phi_{D_n}(R, 0;\hat \alpha) d\hat\alpha.
\end{eqnarray}
The representation of the potential $\Phi_{D_n}$ for $R\ge\alpha$ can be defined by
\begin{eqnarray}
\label{eqnPhiDn}
\Phi_{D_n} = -\frac{\pi G\sigma_0}{2\alpha^{2n-1}}\prod^{n-1}_{j=0}(2j+1)R^{2n} 
\left[ \sum^n_{k=0} b_{n,k} (\xi^2-1)^k \sin^{-1}\xi +\xi \sum^{n-1}_{k=0} c_{n,k} (\xi^2-1)^k \sqrt{1-\xi^2}\right],
\end{eqnarray}
where the coefficients $b_{n,k}$ and $c_{n,k}$ are defined as 
\begin{eqnarray*}
b_{n+1,k}&=&\frac{b_{n,k-1}}{2k}, \mbox{ for } k=1,2,\ldots,n+1,\\
b_{n+1,0}&=& 
(\frac{b_{n,n}}{2n+2}+c_{n,n-1})H(n,0)
+\sum^{n-1}_{k=1}(\frac{b_{n,k}}{2k+1}+c_{n,k}+c_{n,k-1})H(k,0)
+(\frac{b_{n,0}}{4}+\frac{c_{n,0}}{2}),\\
c_{n+1,n}&=& (\frac{b_{n,n}}{2n+2}+c_{n,n-1})\frac{H(n,n)}{2n+1},\\
c_{n+1,m}&=& (\frac{b_{n,n}}{2n+2}+c_{n,n-1})\frac{H(n,m)}{2m+1}
+\sum^{n-1}_{k=m}(\frac{b_{n,k}}{2k+2}+c_{n,k}+c_{n,k-1})\frac{H(k,m)}{2m+1},
\end{eqnarray*}
where $m=1,2,\ldots,n-1$ and 
\begin{eqnarray*}
c_{n+1,0} = (\frac{b_{n,n}}{2n+2}+c_{n,n-1})H(n,0)
+ 
\sum^{n-1}_{k=1}
(\frac{b_{n,k}}{2k+2}+c_{n,k}+c_{n,k-1})H(k,0)
+ 
(\frac{b_{n,0}}{4}+\frac{c_{n,0}}{2}),
\end{eqnarray*}
with initial data $b_{1,0}=1$, $b_{1,1}=2$, and $c_{1,0}=1$, and 
\begin{eqnarray*}
\Phi_{D_n} = -\frac{\pi^2 G \sigma_0}{4\alpha^{2n-1}} 
\prod^{n-1}_{j=0}(2j+1)R^{2n}  \sum^n_{k=0} b_{n,k} (\xi^2-1)^k 
\end{eqnarray*}
for $R\le \alpha$. 
The derivation is straight forward from (\ref{eqnPhiDnp1}) and (\ref{eqnPhiDn}) with the help of the following identities,
\begin{eqnarray*}
\int \xi (\xi^2-1)^k \sin^{-1}\xi d\xi 
&=& \frac{1}{2k+2}(\xi^2-1)^{k+1}\sin^{-1}\xi +\frac{1}{2k+2} \int (\xi^2-1)^k \sqrt{1-\xi^2}d\xi,\\
\int (\xi^2-1)^k \sqrt{1-\xi^2} d\xi
&=& \sum^k_{m=0}\frac{H(k,m)}{2m+1} \xi(\xi^2-1)^m\sqrt{1-\xi^2}+H(k,0) \sin^{-1}\xi,\\
\int \sqrt{1-\xi^2}d\xi
&=& \frac{1}{2}\xi \sqrt{1-\xi^2} +\frac{1}{2}\sin^{-1}\xi.
\end{eqnarray*}
\clearpage

\begin{table}
\begin{center}
\begin{tabular}{|c|c|c|c||c|c|c|}  \hline
$N$ & $E^1_x$   & $E^2_x$    & $E^\infty_x$ & $E^1_R$   &  $E^2_R$  & $L^\infty_R$ \\ \hline 
32	&1.391E-3&2.232E-3&1.253E-2&2.198E-3&3.005E-3&1.253E-2 \\ \hline
64	&3.567E-4&5.292E-4&3.366E-3&5.668E-4&7.221E-4&3.359E-3 \\ \hline
128	&9.022E-5&1.275E-4&8.926E-4&1.435E-4&1.754E-4&8.914E-4 \\ \hline
256	&2.268E-5&3.114E-5&2.358E-4&3.606E-5&4.302E-5&2.355E-4 \\ \hline
512	&5.683E-6&7.675E-6&6.212E-5&9.037E-6&1.063E-5&6.202E-5 \\ \hline
1024&1.423E-6&1.903E-6&1.633E-5&2.262E-6&2.639E-6&1.630E-5 \\ \hline \hline
$N$ &   $O^1_x$   & $O^2_x$   & $O^\infty_x$  & $O^1_R$   &  $O^2_R$  & $O^\infty_R$ \\ \hline
32/64	&1.96&2.08&1.90 &1.95&2.06&1.90 \\ \hline
64/128	&1.98&2.05&1.91	&1.98&2.04&1.91 \\ \hline
128/256	&1.99&2.03&1.92	&1.99&2.03&1.92 \\ \hline
256/512	&2.00&2.02&1.92	&2.00&2.02&1.92 \\ \hline
512/1024&2.00&2.01&1.93	&2.00&2.01&1.93 \\ \hline
\end{tabular}
\end{center}
\caption{\rt{Table demonstrating} 
the errors and order accuracy for the $\sigma_{D_5}$ disk for various number of zones $N=2^k$ 
of the finer grids from $k=5$ to $10$ for the two grid level simulation.
It shows that the order for the $\sigma_{D_5}$ disk is almost second order for each norm.
}
\label{tblExample1}
\end{table}

\begin{table}
\begin{center}
\begin{tabular}{|c|c|c|c||c|c|c|}  \hline
$N$ & $E^1_x$   & $E^2_x$   & $E^\infty_x$  & $E^1_R$   &  $E^2_R$  & $L^\infty_R$ \\ \hline 
32	&8.632E-4&1.600E-3&1.194E-2&1.311E-3&2.109E-3&1.194E-2 \\ \hline
64	&2.206E-4&3.666E-4&3.226E-3&3.393E-4&4.902E-4&3.218E-3 \\ \hline
128	&5.552E-5&8.612E-5&8.581E-4&8.570E-5&1.162E-4&8.567E-4 \\ \hline
256	&1.391E-5&2.067E-5&2.272E-4&2.149E-5&2.806E-5&2.269E-4 \\ \hline
512	&3.481E-6&5.038E-6&6.000E-5&5.380E-6&6.862E-6&5.989E-5 \\ \hline
1024&8.707E-7&1.241E-6&1.580E-5&1.346E-6&1.693E-6&1.577E-5 \\ \hline \hline
$N$ &   $O^1_x$   & $O^2_x$   & $O^\infty_x$  & $O^1_R$   &  $O^2_R$  & $O^\infty_R$ \\ \hline
32/64	&1.97&2.13&1.89&1.95&2.11&1.89 \\ \hline
64/128	&1.99&2.09&1.91&1.99&2.08&1.91 \\ \hline
128/256	&2.00&2.06&1.92&2.00&2.05&1.92 \\ \hline
256/512	&2.00&2.04&1.92&2.00&2.03&1.92 \\ \hline
512/1024&2.00&2.02&1.92&2.00&2.02&1.93 \\ \hline
\end{tabular}
\end{center}
\caption{\rt{Table demonstrating} 
the errors and order accuracy for the $\sigma_{D_5}$ disk for various number of zones $N=2^k$ 
of the finer grids from $k=5$ to $10$ for a three grid level simulation.  
It shows that the order for the $\sigma_{D_5}$ disk is almost second order for each norm.
}
\label{tblExample2}
\end{table}

\begin{table}
\begin{center}
\begin{tabular}{|c|c|c|c||c|c|c|}  \hline
$N$ & $E^1_x$   & $E^2_x$   & $E^\infty_x$  & $E^1_y$ & $E^2_y$ & $E^\infty_y$  \\ \hline 
32	&3.200E-3	&7.782E-3	&4.830E-2	    & 3.347E-3 & 7.751E-3 & 4.789E-2    \\ \hline
64	&7.854E-4	&1.707E-3	&1.340E-2	    & 8.235E-4 & 1.702E-3 & 1.330E-2    \\ \hline
128	&1.993E-4	&4.086E-4	&3.647E-3	    & 2.090E-4 & 4.075E-4 & 3.621E-3    \\ \hline
256	&5.065E-5	&9.975E-5	&9.874E-4	    & 5.310E-5 & 9.951E-5 & 9.810E-4    \\ \hline
512	&1.277E-5	&2.454E-5	&2.666E-4	    & 1.338E-5 & 2.449E-5 & 2.650E-4    \\ \hline
1024&3.206E-6	&6.072E-6	&7.170E-5	    & 3.359E-6 & 6.059E-6 & 7.129E-5    \\ \hline \hline
$N$ &   $O^1_x$ & $O^2_x$   & $O^\infty_x$  & $O^1_y$& $O^2_y$& $O^\infty_y$ \\ \hline
32/64	&2.03	&2.19	& 1.85              &2.02    &2.19    &1.85 \\ \hline
64/128	&1.98	&2.06	& 1.88              &1.98    &2.06    &1.88 \\ \hline
128/256	&1.98	&2.03	& 1.88	            &1.98    &2.03    &1.88 \\ \hline
256/512	&1.99	&2.02	& 1.89	            &1.99    &2.02    &1.89 \\ \hline
512/1024&1.99	&2.02	& 1.89	            &1.99	 &2.02	  &1.89 \\ \hline
\end{tabular}
\end{center}
\caption{\rt{Table demonstrating} 
the errors and order accuracy for the non-axisymmetric case, 
which consists of two $\sigma_{D_5}$ disks with $\alpha=\sqrt{2}/4$
and centers at $(-1/2,0)$ and $(1/2,0)$
for various number of zones $N=2^k$ of the finer grids from $k=5$ to $10$ for a two grid level simulation.  
It shows that the order of accuracy is 2 for $L^1$ and $L^2$ norms, while 
1.9 for $L^\infty$ norm.
}
\label{tblExample3}
\end{table}

\begin{table}
\begin{center}
\begin{tabular}{|c|c|c|c||c|c|c||c|c|c|}  \hline
$N$ & $E^1_x$   & $E^2_x$   & $E^\infty_x$  & $E^1_y$ & $E^2_y$ & $E^\infty_y$  \\ \hline 
32	&1.243E-3	&2.186E-3	&1.340E-2	    &1.257E-3  &2.176E-3  &1.330E-2     \\ \hline
64	&3.161E-4	&5.233E-4	&3.647E-3	    &3.202E-4  &5.212E-4  &3.621E-3     \\ \hline
128	&7.780E-5	&1.251E-4   &9.875E-4	    &7.892E-5  &1.247E-4  &9.810E-4     \\ \hline
256	&1.962E-5	&3.079E-5	&2.666E-4	    &1.991E-5  &3.069E-5  &2.650E-4     \\ \hline
512	&4.970E-6	&7.657E-6	&7.170E-5	    &5.039E-6  &7.631E-6  &7.129E-5     \\ \hline
1024&1.245E-6	&1.901E-6	&1.921E-5	    &1.262E-6  &1.895E-6  &1.911E-5     \\ \hline \hline
$N$ &   $O^1_x$ & $O^2_x$& $O^\infty_x$  & $O^1_y$& $O^2_y$& $O^\infty_y$ \\ \hline
32/64	&1.97	&2.06	& 1.88           & 1.97   &2.06    &1.88           \\ \hline
64/128	&2.02	&2.06	& 1.88           & 2.02   &2.06    &1.88           \\ \hline
128/256	&1.99	&2.02	& 1.89           & 1.99   &2.02    &1.89           \\ \hline
256/512	&1.98	&2.01	& 1.89           & 1.98   &2.01    &1.89           \\ \hline
512/1024&2.00	&2.01	& 1.90	         & 2.00   &2.01    &1.90           \\ \hline
\end{tabular}
\end{center}
\caption{\rt{Table demonstrating} 
the errors and order accuracy for the non-axisymmetric case, 
which consists of two $\sigma_{D_5}$ disks with $\alpha=\sqrt{2}/4$
for various number of zones $N=2^k$ of the finer grids from $k=5$ to $10$ for two grid level simulation.  
It shows that the order of accuracy is 2 for $L^1$ and $L^2$ norms, while 
1.9 for $L^\infty$ norm.
}
\label{tblExample4}
\end{table}

\begin{table}
\begin{center}
\begin{tabular}{|c|c|c|c||c|c|c|}  \hline
$N$ & $E^1_x$   & $E^2_x$    & $E^\infty_x$ & $E^1_R$   &  $E^2_R$  & $L^\infty_R$ \\ \hline 
32	 &1.841E-3&3.495E-3&3.003E-2&2.799E-3&4.718E-3&2.847E-2 \\ \hline
64	 &4.558E-4&9.025E-4&9.752E-3&6.967E-4&1.228E-3&9.179E-3 \\ \hline
128	 &1.208E-4&2.528E-4&3.762E-3&1.848E-4&3.468E-4&3.756E-3\\ \hline
256	 &3.056E-5&6.321E-5&1.272E-3&4.661E-5&8.672E-5&1.250E-3 \\ \hline
512	 &7.786E-6&1.705E-5&4.434E-4&1.187E-5&2.348E-5&4.409E-4 \\ \hline
1024&1.951E-6&4.450E-6&1.531E-4&2.973E-6&6.143E-6&1.516E-4 \\ \hline \hline
$N$ &   $O^1_x$   & $O^2_x$   & $O^\infty_x$  & $O^1_R$   &  $O^2_R$  & $O^\infty_R$ \\ \hline
32/64	     &1.99&1.95&1.62&2.00&1.94& 1.63\\ \hline
64/128	 &1.92&1.84&1.37&1.92&1.82& 1.29\\ \hline
128/256	 &1.98&2.00&1.56&1.99&2.00& 1.59\\ \hline
256/512	 &1.97&1.89&1.52&1.97&1.89& 1.50\\ \hline
512/1024&2.00&1.94&1.53&2.00&1.93& 1.54\\ \hline
\end{tabular}
\end{center}
\caption{ \rt{Table demonstrating} \hhw{the errors and order accuracy for the $\sigma_{D_2}$ disk for 
various number of zones $N=2^k$ 
of the finer grids from $k=5$ to $10$ for the two grid level simulation.
It shows that the order for the $\sigma_{D_2}$ disk is almost second order in terms of $L^1$ and $L^2$ norms, while reduces to $\approx 1.5$ for $L^{\infty}$ norm.}
}
\label{tblD2}
\end{table}

\end{document}